\documentclass[aps,twocolumn,pra]{revtex4-1}
\usepackage{newlfont}
\usepackage{amssymb}
\usepackage{amsfonts}
\usepackage{amsmath}
\usepackage{wasysym}
\usepackage{graphicx}
\usepackage{bm}

\newcommand{\braket}[1]{\langle {#1} \rangle}

\usepackage{epsfig}

\usepackage{amsthm}

\begin{document}

\title{Classical spin models with broken symmetry: Random Field Induced Order and Persistence of spontaneous magnetization in presence of a random field }

\author{Anindita Bera\(^{1,2}\), Debraj Rakshit\(^{2}\), Maciej Lewenstein\(^{3,4}\), Aditi Sen(De)\(^{2}\), Ujjwal Sen\(^{2}\), and Jan Wehr\(^{5}\)}

\affiliation{\(^1\)Department of Applied Mathematics, University of Calcutta, 92, A.P.C. Road, Kolkata 700 009, India\\
\(^2\)Harish-Chandra Research Institute, Chhatnag Road, Jhunsi, Allahabad 211019, India\\
\(^3\)ICREA-Instituci$\acute{o}$ Catalana de Recerca i Estudis Avancats, Lluis Companys 23, 08010 Barcelona, Spain\\
\(^4\)ICFO-Institut de Ci$\grave{e}$ncies Fot$\grave{o}$niques, 08860 Castelldefels (Barcelona), Spain\\
\(^5\)Department of Mathematics, University of Arizona, Tucson, AZ 85721-0089, USA
}


\date{\today}

\begin{abstract}
We consider classical spin models of two- and three-dimensional spins with continuous symmetry and investigate the effect of a symmetry-breaking unidirectional quenched disorder on the magnetization of the system. We work in the mean-field regime. We show, by perturbative calculations and numerical simulations, that although the continuous symmetry of the magnetization is lost due to disorder, the system still magnetizes in specific directions, albeit with a lower value as compared to the case without disorder. The critical temperature, at which the system starts magnetizing, as well as the magnetization at low and high temperature limits, in presence  of disorder, are estimated. Moreover, we treat the $SO(n)$ $n$-component spin model to obtain the generalized expressions for the near-critical scalings, which suggest that the effect of disorder in magnetization increases with increasing dimension. We also study the behavior of magnetization of the classical XY spin model in the presence of a constant 
magnetic field, in addition to the quenched disorder. We find that in presence of the uniform magnetic field, disorder may enhance the component of magnetization in the direction that is transverse to the disorder field.

\end{abstract}


\maketitle

\section{Introduction}
\label{sec_introduction}

Disordered systems lie at the center stage of condensed matter and atomic many-body physics, both classical and quantum~\cite{anderson,ahufinger}. Challenging open questions in disordered systems include those in the realms of spin glasses~\cite{Mezard}, neural networks~\cite{amit}, percolation~\cite{Aharony}, and high $T_c$ superconductivity \cite{Auerbach}. Phenomena like Anderson localization~\cite{anderson2} and absence of magnetization in several classical spin models~\cite{Nagaoka} are effects of disorder.

In particular, classical ferromagnetic spin models with discrete, or continuous, symmetries are very sensitive to random magnetic fields, distributed in accordance with the symmetry, in low dimensions~\cite{imry}. For instance, an arbitrary small random magnetic field with $\mbox{{\bf Z}}_2(\pm)$ symmetry destroys spontaneous magnetization in the Ising model in 2D at any temperature $T$, including $T = 0$. Similar effects hold for the XY model in 2D at $T =0$ in a random field with $U(1)$ $(SO(2))$ symmetry, and Heisenberg model in 2D at $T = 0$ in $SU(2)$ $(SO(3))$ symmetry in random field. In these cases, the effects of disorder amplify the effects of continuous symmetry, that destroys spontaneous magnetization at any $T > 0$. The effect is even more dramatic in 3D, where the random field destroys spontaneous magnetization at any $T \ge 0$ (see \cite{imry,imbrie,Aizenman} for a general description of these).

The appropriate symmetry of the random field is essential for the results mentioned above. The natural question arises as to what happens if the distribution of the random field does not exhibit the symmetry, in particular the continuous symmetry. Yet another natural question is how the spin systems in random fields behave in the quantum limit. The latter question is particularly interesting in view of the fact that nowadays it is possible to realize practically ideal models of quantum spin systems (with spin $s=1/2, 1, 3/2, \cdots,$ and with Ising, XY, or Heisenberg interactions) in controlled random fields~\cite{ahufinger,wehr}. It is therefore very important to understand the physics of both classical and quantum spin models in random fields that break their symmetry.

In this paper, we will consider the classical XY spin model in a random field that breaks its continuous $U(1)$ $(SO(2))$ symmetry. We investigate this model in the mean-field approximation~\cite{Thompson}. Despite its simplicity, this spin model magnetizes in the absence of disorder below a certain critical temperature, which can be calculated exactly. As a result of continuous symmetry, the spontaneous magnetization can have an arbitrary direction. Subsequently, a unidirectional random magnetic field is introduced, by adding a new term to the energy of the model. This term breaks the continuous symmetry of the model, but the critical temperature persists. We find that the system possesses magnetization in specific directions, viz. the direction transverse to that of the random field and along the direction of the random field. The present paper employs numerical as well as perturbative techniques to study the critical behaviour and properties of the magnetization for both cases within a mean field 
framework. We prove that, as may be expected, by adding a random field, the critical temperature in both cases (parallel as well as transverse magnetization) decreases with the increase of the random field strength.  We also show that although the magnetization of the disordered system is lower than that of the pure system (i.e. the system without disorder) near the critical point for both cases, the disorder effect is more pronounced along the direction of the disorder field in this regime.  We work through the low-temperature aspects of the magnetization for the two-dimensional spin system as well. 

Next, we introduce a constant magnetic field, which breaks the continuous symmetry of the model even in the absence of disorder. In fact, the system now magnetizes at all temperatures in the direction parallel to the magnetic field. When we also add a random field (in the $Y$-direction) as described above, the length of the magnetization vector decreases again. Moreover, at low temperature the magnetization gets atrracted towards the direction that is transverse to that of the random field. However, the $X$-component of the magnetization can increase for certain choices of the magnetic field. We view this effect as a ``random field induced order", by analogy of the effect studied in \cite{wehr}, where numerical evidence was given for appearance of magnetization in the XY model on a 
two-dimensional lattice with the introduction of the disorder. In contrast to the present work, in this other case, no mean-field approximation was used and no uniform magnetic field was introduced.

The effect of random field induced order has, of course, a long
history \cite{Aharnoy2}. Recently it has become vividly discussed
in the context of XY ordering in a graphene quantum Hall
ferromagnet \cite{Abanin}, and ordering in $^3He-A$ aerogel and
amorphous ferromagnets \cite{Fomin}. Volovik \cite{Volovik-late}
considered it in the context of the so-called Larkin-Imry-Ma
state  in $^3He-A$ aerogel.  The earlier paper \cite{wehr} clarifies certain
aspects of the rigorous proof of the appearance of magnetization in
XY model at $T = 0$, presentation of a novel evidence for the same
effect at $T > 0$, and a proposal for realization of quantum
version of the effect with ultracold atoms. In the subsequent
paper \cite{Niederberger}, they have shown how the random field
induced order exhibits itself in a system of two-component
trapped Bose-Einstein condensate with random Raman inter-component
coupling. These studies were recently continued in Ref.
\cite{Lellouch}.  Other possible quantum realizations include
disorder-induced phase control in superfluid Fermi-Bose mixtures
\cite{Armand-BCS}, or rounding of   first order transition on low
dimensional quantum  systems \cite{Lebowitz}.

Disorder-induced order persists also in 1D quantum spin chains
\cite{Armand-chain}; the corresponding quantum phase transition is
related to the one occurring for ferromagnetic chains in the
staggered magnetic field (for recent studies, including effects in
spin dynamics,  see \cite{Huirong}). It is also worth mentioning
that there exists an analog of random field induced order in
temporally and spatially disordered/modulated fields (for the
works on generation of solitons and patterns in non-linear wave
equations, see  \cite{Malomed,Kestutis}). Last, but not least, the
effect was also mentioned in the general context of transport in
disordered ultracold quantum gases \cite{Pezze}, localization of
Bogoliubov modes \cite{Lugan}, and disorder-induced trapping and
Anderson localization in expanding Bose condensates \cite{IO1}.

Classical instances of random field induced order concern,
among others, concentration phase transitions \cite{Morosov}, and
loss and recovery of Gibbsianness for XY models in random fields
\cite{Enter}. Recently, the classical XY model in a weak random
field in the $Y$-direction has been considered in Refs.
\cite{{Crawford}, {Nicholas}}. These works form a breakthrough 
in mathematical analysis of lattice spin models, and in particular
proving that the XY model in such random field orders at non-zero $T$,
confirming the conjecture of Ref. \cite{wehr} -- for details of
the very complex proof, see \cite{cmp}. This remarkable result sheds
new light on the mechanism underlying the random-field-induced order. 
The novelty of the present paper lies in systematic mean field treatment 
of the disordered XY model with particular emphasis on the response to 
the constant magnetic field.

We further investigate the classical Heisenberg spin model in presence of the random field. We find that the quenched magnetization of the classical Heisenberg model in the mean field limit behaves similarly to the classical XY model. Finally, we present general expressions of the critical scalings of magnetization for an  $n$-component classical spin system. Specifically, we find that the magnitude of magnetization due to disorder decreases as the square of strength of randomness in all dimensions.

The remainder of this paper is arranged as follows. Sec.~II reviews the ferromagnetic XY model within the mean field approach. A symmetry-breaking random field is added in Sec.~III and the results of numerical simulations and perturbative calculations on the resulting model are presented. Sec.~IV studies the system in the presence of an additional constant field and, in particular, shows the presence of a random field induced magnetization. In Sec.~V, we discuss the classical Heisenberg model in the mean field approximation. Sec.~VI applies the perturbative treatment to compute the generalized expressions of magnetization near criticality for the $SO(n)$-symmetric $n$-component classical spin model.

\section{Ferromagnetic XY model: Mean field approach}
\label{sec_XY_pure}

Consider a lattice $\mbox{{\bf Z}}^d$ of points with integer coordinates in $d$ dimensions, each site $i$ of which is occupied by	a ``spin", which	is a unit vector $\vec{\sigma}_i = (\cos \theta_i, \sin \theta_i)$ on a two-dimensional plane (called the XY plane). The nearest-neighbor ferromagnetic XY model is defined by the Hamiltonian
\begin{eqnarray}
\label{spin-spin-int}
H_{XY}=-J \sum_{|i-j|=1} \vec{\sigma}_i.\vec{\sigma}_j,
\end{eqnarray}
with a coupling constant $J>0$. 
This model does not have any spontaneous magnetization, 
at any temperature, in one and two dimensions (Mermin-Wagner-Hohenberg theorem~\cite{Mermin}), 
while a nonzero magnetization appears in higher dimensions at sufficiently 
low temperatures~\cite{Justin,Frohlich}.

Let us assume that the total number of spins in our system is $N$. 
In the mean field approximation every spin is assumed to interact 
with all other spins (not just with the nearest neighbors) with the 
same coupling constant $-J$. Therefore, the contribution of the 
spin at $i$ to the total energy of the system equals
\begin{eqnarray}
\label{avg-spin-contribution}
\left(-\frac{J}{N} \sum_{j; j \ne i} \vec{\sigma}_j\right).\vec{\sigma}_i,
\end{eqnarray}
where we have divided the energy term by $N$ in order to preserve its 
order of magnitude. This effective interaction, 
replacing the nearest neighbor interaction in $H_{XY}$, 
is for large $N$ approximately equal to
\begin{eqnarray}
\label{ham_MF}
\frac{1}{N}\left(-J \sum_j \vec{\sigma}_j\right). \vec{\sigma_i}
\nonumber \\
=-J \vec{m}.\vec{\sigma}_i,
\end{eqnarray}
where $\vec{m}=\frac{1}{N}\sum_{j=1}^N \vec{\sigma}_j$. The mean field approximation
consists in treating $\vec{m}$ as a genuine constant vector and 
adjusting it so, that the canonical average of the spin at (any) site $i$
 equals this constant. If the system is in canonical equilibrium at temperature $T$, 
the average value of the spin vector $\vec{\sigma}_i$ is
\begin{equation}
\label{qxpectation_spin}
\braket{\vec{\sigma}_i}  =\frac{\int_{0}^{2 \pi} {\vec{\sigma}} \exp(\beta J \vec{m}.\vec{\sigma}) d\vec{\sigma}}{\int_{0}^{2 \pi}  \exp(\beta J \vec{m}.\vec{\sigma}) d\vec{\sigma}},
\end{equation}
where $\beta=1/(k_B T)$, with $k_B$ being the Boltzman constant. This average is independent of the site $i$. Con- sistency requires that the left hand side (l.h.s.) of the above equation be equal to the magnetization $\vec{m}$. Hence, we obtain the mean field equation 
\begin{equation}
\label{m_pure}
\vec{m}=\frac{\int_{0}^{2 \pi} {\vec{\sigma}} \exp(\beta J \vec{m}.\vec{\sigma}) d\vec{\sigma}}{\int_{0}^{2 \pi}  \exp(\beta J \vec{m}.\vec{\sigma}) d\vec{\sigma}},
\end{equation}
where we have dropped the index $i$.
Eq.~(\ref{m_pure}) reduces to (see Appendix A)
\begin{eqnarray}
\label{mnew_pure}
m = \frac {I_{1} [\beta J m]}{I_{0} [\beta J m]},
\end{eqnarray}
where $I_{n}[x]$ is the modified Bessel function of order $n$ with argument $x$. Here $m=|\vec{m}|$.

\begin{figure}[t]
\vspace*{+.9cm}
\includegraphics[angle=0,width=70mm]{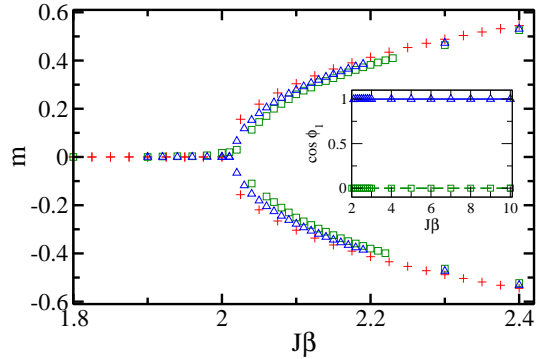}
\vspace*{0.2cm}
\caption{(Color online.) Length of magnetization, $m$, as a function of $J\beta$ for the XY model with the disorder. Pluses represent the solutions for the pure system. Triangles and squares represent respectively the numerical data for transverse (case I) and parallel (case II) magnetization. Inset: Cosine of the angle associated with the magnetization vector $\vec{m}$ as a function of $J\beta$ for the two different cases. Similar symbols as in the main diagram are used in the inset to represent the two different cases. The data, for which $\cos \phi_1 \approx 1$, suggests that the magnetization belongs to case I. Otherwise, it belongs to case  II, where  $\cos \phi_1 \approx 0$. $\phi_1$ is measured in radians. All other axes represent dimensionless quantities.}
\label{fig_effective}
\end{figure}
The red pluses in Fig.~\ref{fig_effective} show the magnetization, $m$, of the pure system as a function of $J \beta$. For sufficiently high temperatures, the only solution for the mean field equation is $\vec{m}=0$. The numerical simulations support the existence of a $\beta_c^{0,2}$, such that for $\beta > \beta_c^{0,2} \approx 2$, 
this system magnetizes. The $1^{st}$ superscript of $\beta_c^{0,2}$ indicates that the system is without disorder and the $2^{nd}$ one denotes the component of the spins. (The superscripts of $\beta_c^{0,2}$ are anticipating  the cases with disorder and higher dimensional spins.) 
By symmetry, the magnetization of the system behaves uniformly in all possible orientations, 
which implies that the solutions of the above mean field equation (Eq.~(\ref{m_pure})) 
form a circle of radius $m^{0,2}$ in the XY plane for a given $J \beta$. Note that the supersripts of $m^{0,2}$ follows the same conventions as explained before for superscripts of the critical temperature.

Approximate analytical expressions for the $\beta_c^{0,2}$ and the behavior of the magnetization $m^{0,2}$ near criticality can be obtained perturbatively. Note that finding magnetization in Eq.~(\ref{mnew_pure}) is equivalent to finding the zeros of the function,
\begin{equation}
\label{f_pure}
F_{2}(m)=\frac {I_{1} [\beta J m]}{I_{o} [\beta J m]}-m.
\end{equation}
If we expand $F_2(m)$ for small $m$, we obtain
\begin{equation}
\label{f_pure_expand}
F_{2}(m)=(-1+\frac{J \beta}{2}) m-\frac{1}{16}(J^3 \beta^3) m^3+o(m^4).
\end{equation}
The nontrivial roots of Eq.~(\ref{f_pure_expand}) are given by
\begin{equation}
\label{sol_m0_d2}
m^{0,2} = \pm \frac{2\sqrt{2}} {J^{3/2}} \beta^{-3/2} (J\beta-2)^{1/2}.
\end{equation}
Therefore, within this approximation, $m^{0,2}$ vanishes if $J \beta=2$, has non-zero values iff  $J \beta>2$, and the critical temperature is given by
\begin{equation}
\label{crit_d2}
\beta_c^{0,2}=\frac{2}{J}.
\end{equation}

\section{Ferromagnetic XY model in a random field}
\label{sec_XY_random}

We now consider the effect of additional quenched random fields in the system. 
Let us begin by the notions of quenched disorder and quenched averaging.

\subsection{Quenched averaging}

The disorder considered in this paper is ``quenched", i.e., its configuration remains unchanged for a time that is much larger than the duration of the dynamics considered. In the systems that we study, it is the local magnetic fields that are disordered. They are random variables with a certain probability distributions. Since the disorder is quenched, a particular realization of all the random variables remains fixed for the whole time necessary for the system to equilibrate. An average of a physical quantity, say $A$, is thus to be carried out in the following order:

(a) Compute the value of the physical quantity A, with the fixed configuration of the disorder.

(b) Average over the disordered parameters.

This mode of averaging is called {\it quenched}. It may be mentioned that an averaging in which items (a) and (b) are interchanged in order, is called {\it annealed} averaging. Physically it corresponds to a situation when the disorder fluctuates on time scales comparable to the system's thermal fluctuations.

\subsection{The model and the mean field equation for magnetization}
The XY model with an inhomogeneous magnetic field has the interaction
\begin{equation}
\label{ham_random-0}
H=-J \sum_{|i-j|=1} \vec{\sigma}_i.\vec{\sigma}_j-\epsilon \sum_i\vec{h}_i.\vec{\sigma}_i,
\end{equation}
where the two-dimensional vectors $\vec{h}_i$ are the external
magnetic fields, up to a coefficient $\epsilon$. In the sequel, $\vec{h}_i$, which 
are random variables of order one, model the disorder in
the system and thus $\epsilon$ measures the disorder strength.
More precisely, let $\vec{h}_i$ be independent and identically distributed random variables (vector-valued). We want to study the effect of including such a random field term in the XY Hamiltonian at small values of $\epsilon$. As argued in \cite{wehr}, in lattice XY models, this effect depends critically on the properties of the probability distribution of the random fields.

If the distribution of the $\vec{h}_i$ is invariant under rotations, there is no spontaneous magnetization at any nonzero temperature in any dimension $d \le 4$ \cite{imry, imbrie, Aizenman}. 

We now want to see the effect of a random field that does not have the rotational 
symmetry of the XY model interaction, by considering the case when
\begin{equation}
\label{hi}
\vec{h}_i=\eta_i.\widehat{e}_y,
\end{equation}
where $\eta_i$ are scalar random variables with a distribution symmetric about 0 and $\widehat{e}_y$ denoting the unit vector in the $y$ direction. The main result of \cite{wehr} is that on the two-dimensional lattice such a random field will break the continuous symmetry and the system will magnetize, even in two dimensions, thus destroying the Mermin- Wagner-Hohenberg effect. Above two dimensions, the pure XY model magnetizes at low temperatures and it has been suggested in \cite{wehr} that the uniaxial random field as described above may enhance this magnetization. In the present paper we want to study related effects at the level of a simpler, mean-field model, which allows for a more detailed analysis and more accurate simulations.

The mean-field Hamiltonian in this case is given by
\begin{equation}
\label{ham_random}
H=-J \vec{m}.\vec{\sigma}-\epsilon \vec{\eta}.\vec{\sigma},
\end{equation}
where $\vec{\eta}$ is the quenched random field in the y-direction, $\vec{\eta}=\eta.\hat{e}_{y}$.
The random variable here is assumed to be Gaussian with zero mean and unit variance. $\epsilon (>0)$ is typically a small parameter that quantifies the strength of randomness. In the mean field equation, the magnetization, which is obtained by averaging over the disorder, is given by
\begin{equation}
\label{m_random}
\vec{m}=Av_{\eta}\left[\frac{\int_{0}^{2 \pi} {\vec{\sigma}} \exp(\beta J \vec{m}.\vec{\sigma}+\beta \epsilon \eta \sigma_{y}) d\vec{\sigma}}{\int_{0}^{2 \pi} 
\exp(\beta J \vec{m}.\vec{\sigma}+\beta \epsilon \eta \sigma_{y}) d\vec{\sigma}}\right],
\end{equation}
where $Av_\eta(\cdot)$ denotes the average over the disorder, i.e., the integral over $\eta$ with the appropriate distribution (here assumed to be unit normal). Set $\vec{m} = (m \cos \phi_1, m \sin \phi_1)$.
As demonstrated in Appendix A, we obtain a coupled set of the following two equations:
\begin{equation}
\label{mx_random_analytical}
m^{\epsilon,2}_{\perp} \equiv m \cos \phi_1=Av_{\eta}\left[\cos \alpha \frac {I_1[\beta r]}{I_0[\beta r]}\right],
\end{equation}
and
\begin{equation}
\label{my_random_analytical}
m^{\epsilon,2}_{\parallel} \equiv m \sin \phi_1=Av_{\eta}\left[\sin \alpha \frac {I_1[\beta r]}{I_0[\beta r]}\right],
\end{equation}
where
\begin{equation}
\label{def_r}
r=\beta \sqrt{J^2 m^2+\epsilon^2 \eta^2+2 Jm \epsilon \eta \sin \phi_1},
\end{equation}
and 
\begin{equation}
\label{def_alph}
\alpha=\text{arctan}\left[ \frac{J m \sin \phi_1+\epsilon\eta}{J m \cos \phi_1}\right].
\end{equation}
and $I_0, I_1$ denote Bessel functions.
\subsection{Contour analysis: Departure from isotropy}
\begin{figure}[t]
\vspace*{+.4cm}
\includegraphics[angle=0,width=70mm]{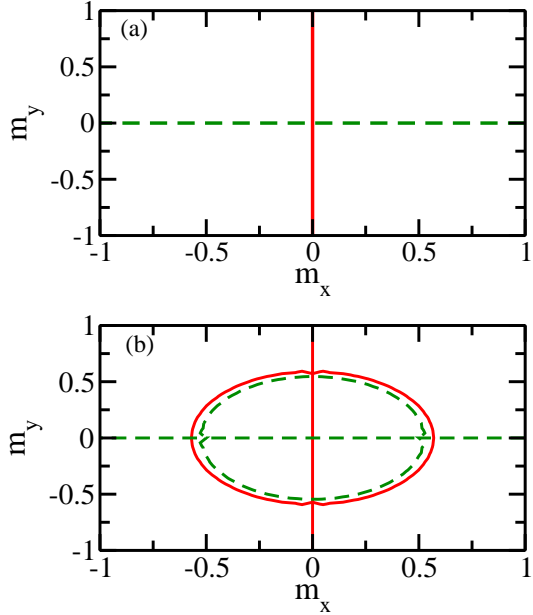}
\vspace*{0.2cm}
\caption{(Color online.) Zero contour lines of the $F_{2,x}^{\epsilon}(m)$ and $F_{2,y}^{\epsilon}(m)$ given in Eqs.~(\ref{fx_approx}) [solid-red] and (\ref{fy_approx}) [dotted-green] for  $\epsilon/J= 0.1$ and $J\beta$ =1.5  (a) and  2.5  (b), respectively, as  functions of $m_x$ and $m_y$. All quantities are dimensionless.}
\label{contour}
\end{figure}
In order to find the magnetization $\vec{m}$ we have to solve the coupled set of Eqs.~(\ref{mx_random_analytical}) and (\ref{my_random_analytical}), which is equivalent to finding the common zeros of the following two functions: 
\begin{equation}
\label{fx}
F_{x}^{\epsilon,2}(m)=Av_{\eta}\left[\cos \alpha~ \frac {I_1[r]}{I_0[r]}\right]-m \cos \phi_1,
\end{equation}
and
\begin{equation}
\label{fy}
F_{y}^{\epsilon,2}(m)=Av_{\eta}\left[\sin \alpha \frac {I_1[r]}{I_0[r]}\right]-m\sin \phi_1.
\end{equation}

Before discussing the numerical results, we first examine the behavior of magnetization for small $\epsilon$ by using a contour analysis within the perturbative framework, providing qualitative insight about the system's behavior.  We perform a Taylor series expansion of the functions given in Eqs.~(\ref{fx}) and (\ref{fy}), in $\epsilon$ around $\epsilon=0$ and obtain
\begin{eqnarray}
\label{fx_approx}
F_{x}^{\epsilon,2}(m)=c_1+b_1\epsilon^{2}+ o(\epsilon^3),
\end{eqnarray}
and
\begin{eqnarray}
\label{fy_approx}
F_{y}^{\epsilon,2}(m)
=c_2+b_2\epsilon^{2}+ o(\epsilon^3).
\end{eqnarray} 
\begin{widetext}
The expansion coefficients $c_i$'s and $b_i$'s are defined as
\begin{eqnarray}
\label{c1}
c_1 = m_x \left(-1 + \frac{ I_1[J m \beta]} {m I_0[J m \beta]}\right),
\end{eqnarray}
\begin{eqnarray}
\label{b1}
b_1 = \bigl(m_x \beta  \bigl(2 J \beta m m_y^2 I_1[J m \beta]^3
+(m_x^2-3 m_y^2) I_0[J m \beta]^2I_2[J m \beta]
\nonumber\\
- I_0[J m \beta]I_1[J m \beta]
\left((m_x^2-m_y^2)I_1[J m \beta]+2 J m \beta m_y^2 I_2[J m \beta]\right) \bigr)\bigr)/
\left(2 J m^4 I_0[J m \beta]^3\right),
\end{eqnarray}
\begin{eqnarray}
\label{c2}
c_2 = m_y \left(-1 + \frac{ I_1[J m \beta]} {m I_0[J m \beta]}\right),
\end{eqnarray}
and
\begin{eqnarray}
\label{b2}
b_2 =-\bigl(m_y \beta \bigl(-2 J^2 \beta m^2 m_y^2 I_1[J m \beta]^3 
-J m (3 m_x^2-m_y^2) I_0[J m \beta]^2 I_2[J m \beta]
\nonumber\\
+I_0[J m \beta] I_1[J m \beta] 
\bigl( J \beta m (3 m_x^2+m_y^2)  I_1[J m \beta]+2 \beta J^2 m^2m_y^2 I_2[J m \beta]\bigr)\bigr)\bigr)/
\left(2 J^2 m^5 I_0[J m \beta]^3\right),  
\end{eqnarray}
where $m_x=m \cos \phi_1, m_y=m \sin \phi_1$.
\end{widetext}

Each of the functions, $F_{x}^{\epsilon,2}(m)$ and $F_{y}^{\epsilon,2}(m)$,
has zero and nonzero contour lines. The zero contour lines are of interest to us. The roots are those that are common to the zero contours of $F_{x}^{\epsilon,2}(m)$ and $F_{y}^{\epsilon,2}(m)$. Figure \ref{contour} shows the contour plots (only the zero contour lines) of  $F_{x}^{\epsilon,2}(m)$ and $F_{x}^{\epsilon,2}(m)$, given in Eq.~(\ref{fx_approx}) and Eq.~(\ref{fy_approx}), respectively, for $\epsilon/J= 0.1$ and for $J\beta = 1.5$ (Fig.~2(a)) and  $J\beta = 2.5$ (Fig.~2(b)), as functions of $m_x$ and $m_y$. For $J\beta=1.5$, the only solution is $m_x=m_y=0$, which signifies the absence of the magnetization in the system below a certain critical temperature. As seen in Fig.~2(b), we have nontrivial solutions for $J\beta$=2.5. The disorder, however, breaks the isotropic symmetry and the  solutions of the  $F_{x}^{\epsilon,2}(m)$ and $F_{y}^{\epsilon,2}(m)$ exist only at $\phi_1=0$ or $\pi/2$. This implies that the system magnetizes either along the transverse direction of the disorder field (case I) 
or along the direction of the disorder field (case II). Note that for $\epsilon=0$, the zero contour lines, i.e., the  solid and dashed lines in Fig.~\ref{contour} (b), would coincide, impling uniformity in magnetization in all possible directions. An arbitrarily small disorder, however, sets the contour lines apart. The contour analysis indicates that there is also a critical temperature in the system below which the system magnetizes, albeit in a different way than in the case without disorder.

\begin{figure}[t]
\vspace*{+.4cm}
\includegraphics[angle=0,width=70mm]{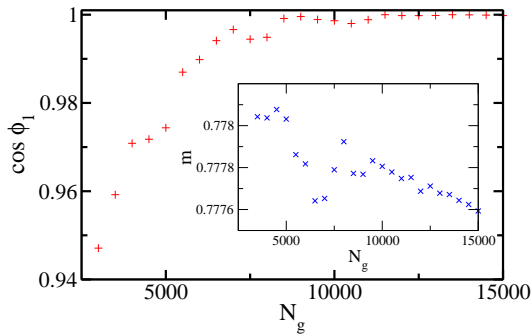}
\vspace*{0.2cm}
\caption{(Color online.) Illustration of convergence during quenched averaging. Pluses represent the cosine of the phase $\phi_1$ as a function of $N_g$, which is the number of the Gaussian distributed random $\eta$ values with disorder strength $\epsilon/J=0.15$ at $J \beta=3.5$. Inset: Crosses show the length of magnetization $m$ as a function of $N_g$ for the same parameters. $\phi_1$ is measured in radians, and $N_g$ in the number of random points generated. All other quantities are dimensionless.}
\label{convergence}
\end{figure}
\subsection{Numerical simulations}
As discussed in the previous section, both transverse magnetization and parallel magnetization survive below some critical temperature. Before analyzing their behavior, let us first explain their quenching mechanism that has been carried out. To find the roots of the Eqs.~(\ref{mx_random_analytical}) and (\ref{my_random_analytical}) for a given $\epsilon$ and $\beta$, we use classical Monte-Carlo technique for performing averaging over $\eta$. We  test convergence of the solutions of the averaged equations as the number of  Gaussian distributed random numbers, $N_g$ increases. We find that it typically requires a few thousand of random numbers to reach the desired convergence. Fig.~\ref{convergence} shows an example of convergence of the phase of $\vec{m}$ for the system with $\epsilon/J=0.15$ and $J \beta=3.5$. The pluses correspond to the cosine of the phase of the magnetization. We find that $\cos \phi_1$ converges to unity for this case (the transverse magnetization). The inset of Fig.~\ref{convergence} 
shows the length of the magnetization $m$, which has converged up to the third decimal point for $N_g> 5000$. As discussed in Section C, there is a second kind of solution for which $\cos \phi_1$ would converge to zero, implying that the system magnetizes along the $y$-axis, i.e.,  along the direction of the disorder field. Below we briefly present the results obtained by numerical simulations for these two different cases.

\subsubsection{Case I: Transverse magnetization}
\begin{figure}[h]
\vspace*{+.2cm}
\includegraphics[angle=0,width=70mm]{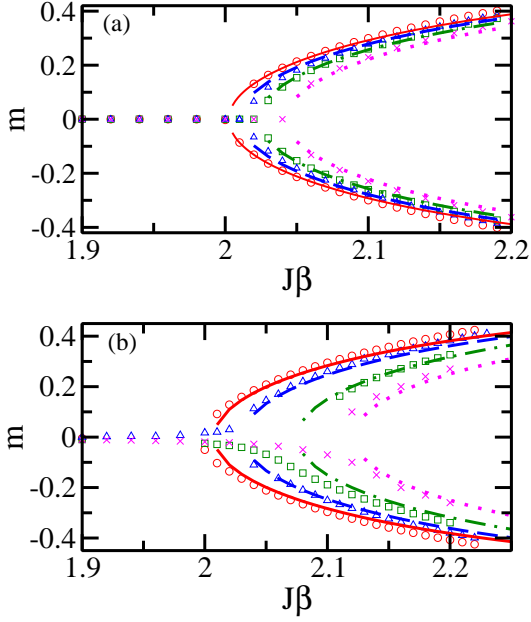}
\vspace*{0.2cm}
\caption{(Color online.) Magnetization as a function of $J \beta$ for (a) transverse direction (case I) and (b) parallel direction (case II) of the disorder field. Circles, triangles, squares and crosses correspond to the solutions of the Eqs.~(\ref{mx_random_analytical}) and (\ref{my_random_analytical}) with $\epsilon/J$ = 0.05, 0.1, 0.15 and 0.2, respectively. The lines in (a) and (b) correspond to the analytical solutions derived for small $m$ given in Eq.~(\ref{sol_m1}) and Eq.~(\ref{sol_m2}), respectively, except the case, where $\epsilon/J$ = 0.2 in (b), for which we had to consider the next higher order contribution in $\epsilon$ in order to achieve good agreement with the numerics.  The expressions are long and we do not include them. All quantities are dimensionless.}
\label{r2_sm}
\end{figure}

As argued by using the contour diagram (Fig.~2), either the $Y$-component or the $X$-component of the magnetization vanishes. Let us discuss the case when $m^{\epsilon,2}_{\perp} \ne 0$, $m^{\epsilon,2}_{\parallel} = 0$. In the case, when $\epsilon \ne 0$, the system again does not magnetize at high temperature  (as in the case of $\epsilon=0$). However, there exists a critical temperature, below which a transverse (with respect to the direction of the random field) magnetization appears. More precisely, there exists a $\beta_{c,\perp}^{\epsilon,2}$ such that for $\beta > \beta_{c,\perp}^{\epsilon,2}$, the magnetization equations have two solutions with vanishing $Y$-components and non-zero $X$-components, having magnitude $\pm m$ (along with the trivial solution $m_x = 0, m_y = 0$). Here $m=|\vec{m}|$. We investigate the dependence of $m$ on the temperature and on the disorder strength $\epsilon$ (see Fig.~\ref{r2_sm} (a)). All the curves show two real solutions ($m_x$ and $-m_x$ in this case) of the 
corresponding mean field equations ~(\ref{mx_random_analytical}) and (\ref{my_random_analytical}). We find that the critical point $\beta_{c,\perp}^{\epsilon,2}$ shifts towards a higher value with increasing $\epsilon$, which implies a lowering of the critical temperature with increasing disorder strength. The scaling of magnetization near the critical point and the low-temperature behavior of the magnetization will be discussed in the following section. 

\subsubsection{Case II: Parallel magnetization}
In this case, the spontaneous magnetization has an approximately zero $X$-component and a nonzero $Y$-component, equal $\pm m$. There is no  magnetization at very high temperature and only below a critical temperature, $\beta_{c,\parallel}^{\epsilon,2}$, the magnetization, which is oriented parallel to the direction of the disorder field, appears in the system. Fig.~\ref{r2_sm} (b) shows the dependence of $m$ on the temperature and on the disorder strength $\epsilon$.  The circles, triangles, squares and crosses represent the cases with $\epsilon/J$ = 0.05, 0.1, 0.15 and 0.2, respectively. We find that the critical temperature, $\beta_{c,\parallel}^{\epsilon,2}$, shifts towards an even higher value compared to the case with $m_x \ne 0$, with increasing $\epsilon$. It appears that the effect of the disorder is more pronounced in the direction of the disorder field than in the transverse direction. This effect remains true for the nontrivial solutions in the high temperature regime, as for a given $\beta$, the 
magnitude of magnetization in the transverse direction is lower than that in the parallel direction (see Fig.~\ref{r2_sm}).  We also find that the magnetization for this case has markedly different low-temperature behavior (shown in Fig.~\ref{large_beta}) than in the previous case. The details will be spelled out in Sec.~III E2.

\subsection{Scaling of critical temperature and magnetization with disorder: Perturbative approach}

We now adopt a perturbative approach for the mean field Hamiltonian in Eq.~(\ref{m_random}) to derive analytical expressions for characterizing the small $m$ behavior in the system and also to obtain expressions for the magnetization at very low temperature. The analytical results are compared with the numerical data obtained in the last subsection.
\subsubsection{Critical point and scaling of magnetization near criticality}
We start with Eqs.~(\ref{fx_approx}) and (\ref{fy_approx}) and perform Taylor expansions around $m=0$. We obtain:
\begin{eqnarray}
\label{fx_sm}
F_{x}^{\epsilon,2}(m) = -\frac{1}{16}\left((16+J\beta(-8+\beta^2 \epsilon^2)) \cos \phi_1\right) m 
\nonumber\\
 -\frac{1}{48} \left(J^3 \beta^3(3-2 \beta^2\epsilon^2+\beta^2\epsilon^2 \cos 2\phi_1) \cos \phi_1\right) m^3
\nonumber\\
+o(m^5),
\end{eqnarray}
and
\begin{eqnarray}
\label{fy_sm}
F_{y}^{\epsilon,2}(m)=-\frac{1}{16}\left((16+J\beta(-8+\beta^2 \epsilon^2)) \sin \phi_1\right) m
\nonumber\\
 -\frac{1}{48} \left(J^3 \beta^3(3-4 \beta^2\epsilon^2+\beta^2\epsilon^2 \cos 2 \phi_1) \sin \phi_1\right) m^3
\nonumber\\
+o(m^5).
\end{eqnarray}
As we have discussed above, the allowed values of $\phi_1$ are $\pi/2$ (system magnetizes in direction parallel to disordered field) and $0$ (system magnetizes in direction transverse to disordered field). For transverse magnetization, $m_{\perp}^{\epsilon,2}$,$F_{y}^{\epsilon,2}(m)$ vanishes and Eq.~(\ref{fx_sm}) has two nontrivial solutions:
\begin{eqnarray}
\label{sol_m1}
m_{\perp}^{\epsilon,2} = \pm \sqrt{3} \sqrt{ \frac{
 16 - 8 J \beta + J \beta^3 \epsilon^2}{-3 J^3 \beta^3 +
  J^3 \beta^5 \epsilon^2}}
\end{eqnarray}
\begin{eqnarray}
\label{sol_m1_1}
\approx \pm m^{0,2} (1 \mp \frac{\beta^2}{8(J \beta-2)} \epsilon^2),
\end{eqnarray}
where $m^{0,2}$ is given by Eq.~(\ref{sol_m0_d2}).
Note that we use the $\perp$ subscript in $m_{\perp}^{\epsilon,2}$ to distinguish it from the parallel magnetization, which is denoted by $m_{\parallel}^{\epsilon,2}$. Similar convention will be followed for the critical temperature.
The critical  point can be obtained by setting $m_{\perp}^{\epsilon,2}=0$ in Eq.~(\ref{sol_m1}) and we get
\begin{eqnarray}
\label{crit_my_con}
16 - 8 J \beta_{c,\perp}^{\epsilon,2} + J (\beta_{c,\perp}^{\epsilon,2})^3 \epsilon^2=0,
\end{eqnarray}
which gives
\begin{eqnarray}
\label{crit_my}
\beta_{c,\perp}^{\epsilon,2} \approx \beta_c^{0,2}+\frac{\epsilon^2}{ J^3}.
\end{eqnarray}
Therefore, we obtain corrections of order $\epsilon^2$  to the critical temperature, as observed also in the numerical simulations (see Fig.~\ref{r2_sm}).
A comparison of the analytical expressions (Eq.~\ref{sol_m1}) with the numerical results for the case I with small $m_{\perp}^{\epsilon,2}$ has been made in Fig.~\ref{r2_sm}(a). It is clear that the  results are in good agreement for small $m_{\perp}^{\epsilon,2}$ but not for large $m_{\perp}^{\epsilon,2}$.

Next we find out the expressions for case II by inserting $\phi_1=  \pi/2$ in Eqs.~(\ref{fx_sm}-\ref{fy_sm}). In this case, Eq.~(\ref{fx_sm}) vanishes and Eq.~(\ref{fy_sm}) has two nontrivial solutions:
\begin{eqnarray}
\label{sol_m2}
m_{\parallel}^{\epsilon,2} = \pm \sqrt{3} \sqrt{ \frac{
 16 - 8 J \beta +3 J \beta^3 \epsilon^2}{-3 J^3 \beta^3 +5 J^3 \beta^5 \epsilon^2}},
\end{eqnarray}
which can again be written as
\begin{eqnarray}
\label{sol_m2_1}
m_{\parallel}^{\epsilon,2}  \approx \pm m^{0,2} (1 \mp \frac{3 \beta^2}{8(J \beta-2)} \epsilon^2),
\end{eqnarray}
By setting $m_{\parallel}^{\epsilon,2} = 0$ in Eq.~(\ref{sol_m2}), we obtain the equation for the critical temperature as 
\begin{eqnarray}
\label{crit_my_con1}
16 - 8 J \beta_{c,\parallel}^{\epsilon, 2}  +3  J (\beta_{c,\parallel}^{\epsilon, 2} )^3 \epsilon^2=0,
\end{eqnarray}
which gives
\begin{eqnarray}
\label{crit_my1}
\beta_{c,\parallel}^{\epsilon,2}  \approx \beta_c^0+ \frac{3 \epsilon^2}{ J^3}.
\end{eqnarray}
Therefore, we again obtain $\epsilon^2$ corrections to the critical temperature.
By comparing Eq.~(\ref{sol_m1_1}) and Eq.~(\ref{sol_m2_1}), we note that the $m_{\parallel}^{\epsilon,2} $ is smaller than the $m_{\perp}^{\epsilon,2} $. In Fiq.~\ref{r2_sm}(b), we compare the analytical and numerical results for various disorder strengths.

\subsubsection{Scaling of magnetization at low temperatures}
\label{scaling-low-2d}
We now study the behavior of $m$ at low temperatures, i.e. for large $\beta$. 
We start from the case $\epsilon = 0$. Note that the
numerator and denominator of  Eq.~(\ref{m_pure}) are of the form of $I_n(z)$. Therefore, for
large $\beta$, we use the asymptotics of the Bessel function (see Appendix B) and obtain the following equation for $m$:
\begin{eqnarray}
\label{pure_large_beta}
m^3-m^2+\frac{m}{2\beta J}+o(1/\beta^2)=0.
\end{eqnarray}
Since $m \to \pm 1$ as $\beta \to \infty$, let us write m as
\begin{eqnarray}
\label{pure_large_beta1}
m =\pm 1 \mp \frac{a_1}{\beta}+o(1/\beta^2).
\end{eqnarray}
Putting this in Eq.~(\ref{pure_large_beta}), we finally obtain the behavior of the magnetization for the case when $\epsilon = 0$, for large $\beta$:
\begin{eqnarray}
\label{pure_large_beta2}
m^{0,2} =\pm 1 \mp \frac{1}{2 J \beta}+o(1/\beta^2).
\end{eqnarray}
Using a similar technique for the disordered case, we can perform series expansions for large $\beta$ of Eqs.~(\ref{fx_approx}) and (\ref{fy_approx}). Considering only the leading order contributions from $\epsilon$ and $\beta$, we obtain:
\begin{eqnarray}
\label{fx_large_beta}
F_{x}^{\epsilon,2}(m) = \cos \phi_1 - m \cos \phi_1 -\frac{\cos \phi_1} {2 J m \beta}
\nonumber\\
+\frac{\epsilon^2 \cos \phi_1 (1-3 cos  (2\phi_1))} {4 J^2 m^2}+\cdots
\end{eqnarray}
and
\begin{eqnarray}
\label{fy_large_beta}
F_{y}^{\epsilon,2}(m) = \sin \phi_1-m \sin \phi_1-\frac{\sin \phi_1}{2 J \beta m}
\nonumber\\
-\frac{3 \epsilon^2 \cos^2 \phi_1 \sin \phi_1}{2 J^2 m^2}+\frac{\epsilon^2  \sin \phi_1(1+2 \cos (2\phi_1))}{2 J^3 m^3 \beta} + \cdots \nonumber \\
\end{eqnarray}
\begin{figure}[t]
\vspace*{+.4cm}
\includegraphics[angle=0,width=70mm]{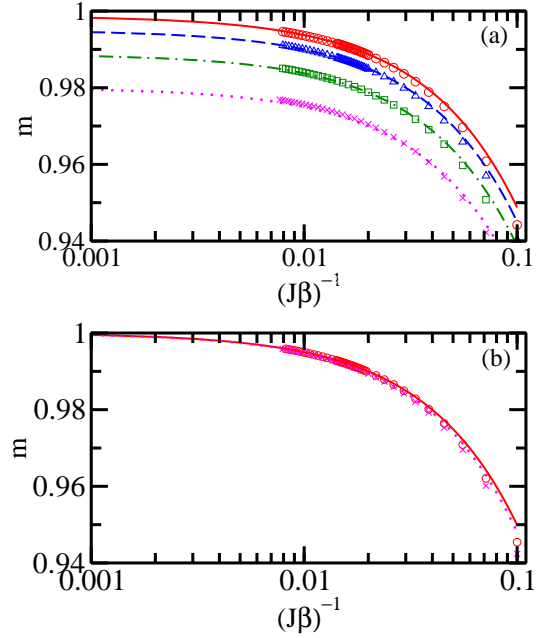}
\vspace*{0.2cm}
\caption{(Color online.) Magnetization as a function of temperature for (a) the transverse direction (case I) and (b) the parallel direction (case II). (a) Circles, triangles, squares and stars correspond to the numerical data for $\epsilon/J$=0.05, 0.1, 0.15 and 0.2, respectively. The solid, dashed, dashed-dotted and dotted lines correspond 
to the analytical solution derived for large $\beta$ given in Eq.~(\ref{m_large_beta_c1}) for the same values of $\epsilon/J$ respectively. (b) Circles and crosses correspond to the numerical data for $\epsilon/J$=0.05 and 0.2, respectively. The solid and the dotted lines correspond to the analytical solution derived for large $J\beta$ given in Eq.~(\ref{m_large_beta_c2}), for the same values of $\epsilon/J$, respectively. In case (b), we have also performed calculations for other small values of $\epsilon/J$ . We are not displaying them here, as they are very close to the displayed ones. All quantities are dimensionless. }
\label{large_beta}
\end{figure}

For the transverse case, where $\phi_1=0$, Eq.~(\ref{fy_large_beta}) vanishes, and from Eq.~(\ref{fx_large_beta}), we obtain
\begin{eqnarray}
\label{m_large_beta_c1}
m_{\perp}^{\epsilon,2} \approx \pm 1 \mp \frac{1}{2 J \beta} \mp \frac{\epsilon^2}{2 J^2}+\cdots
\end{eqnarray}
As $\beta \to \infty$, $m_{\perp}^{\epsilon,2}=1-(\epsilon^2/(2 J^2))$, i.e., the disorder leads to corrections of order $\epsilon^2$ to the magnetization at low temperature. Figure 5 shows the magnetization at low temperature. The circles, triangles, squares and crosses in Fig.~\ref{large_beta}(a) correspond to the data from numerical simulations and the lines correspond to the  analytical expression given in Eq.~(\ref{m_large_beta_c1}).

Now we consider the parallel case, where $\phi_1=\pi/2$, so that  Eq.~(\ref{fx_large_beta}) is automatically satisfied.  The solution to Eq.~(\ref{fy_large_beta}) is given by
\begin{eqnarray}
\label{m_large_beta_c2}
m_{\parallel}^{\epsilon,2} \approx \pm 1 \mp \frac{1}{2 J \beta} \mp\frac{\epsilon^2}{2 J^3 \beta}+\cdots
\end{eqnarray}
For zero temperature, i.e., infinitely large $\beta$, the magnetization of the system reaches unity, which is notably different from that of the previous case, where $\epsilon$ leaves an imprint even at zero temperature. This also implies that even though the disorder had an effect at small $m$ along the direction of the field, the effect is eventually nullified at sufficiently low temperature. The circles and the crosses in Fig.~\ref{large_beta}(b) correspond to the data from numerical simulations and the lines correspond to the analytical expression (Eq.~(\ref{m_large_beta_c2})).

\section{FERROMAGNETIC XY MODEL IN A RANDOM FIELD PLUS A CONSTANT FIELD: RANDOM FIELD INDUCED ORDER}
\label{sec_XY_steady_random}
We have seen in the preceding section that a random field that breaks the symmetry of the XY model, restricts possible magnetization values to a discrete set. Although the system still magnetizes, we no longer have continuous symmetry of the set of solutions to the mean field equation, as a result of adding a symmetry-breaking random field. In this section we explore the effects of such a random field on a system which already has a unique direction of the magnetization, determined by a uniform magnetic field.

\begin{figure}[]
\vspace*{+.4cm}
\includegraphics[angle=0,width=70mm]{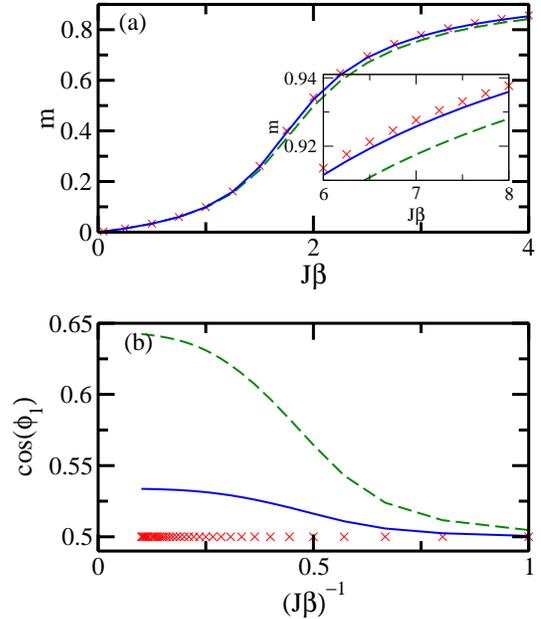}
\vspace*{0.2cm}
\caption{(Color online.) (a) Magnetization, $m$, as a function of $J \beta$. Red crosses represent the case when the XY model has the applied constant field $\vec{h}$ with $h/J=0.1$ and $x = \pi/3$. The blue solid and the green dashed lines show $m$ of the system with the additional random field of strength $\epsilon = 0.1 J$ and $\epsilon = 0.2 J$, respectively. (b) Cosine of phase of magnetization, $\cos \phi_1$, as a function of $1/(J \beta)$ for the same system. $\phi_1$ and $x$ are measured in radians. All other quantities are dimensionless.} 
\label{mag-consth}
\end{figure}

First consider the case in which the planar symmetry in the XY model is broken by applying a constant magnetic field $\vec{h}$ alone. That is, according to the general mean field strategy, we are looking for the solutions of the following equation:
\begin{eqnarray}
\label{m_h}
\vec{m}=\frac{\int_{0}^{2 \pi} {\vec{\sigma}} \exp(\beta J \vec{m}.\vec{\sigma}+\beta \vec{h}.\vec{\sigma}) d\vec{\sigma}}{\int_{0}^{2 \pi}  \exp(\beta J \vec{m}.\vec{\sigma}+\beta \vec{h}.\vec{\sigma}) d\vec{\sigma}},
\end{eqnarray}
Let $\vec{h} = (h \cos x, h \sin x)$. We assume that $0 < h \le 1$ and $-\pi/2 \le x \le \pi/2$. As expected, due to the applied constant field, the mean field equation has a unique solution of the magnetization $\vec{m}$ at all temperatures,  and the solution is a (positive) multiple of $\vec{h}$, but of reduced magnitude. We sum up the situation in Fig.~\ref{mag-consth}. Red crosses correspond to the magnitude, $m$ (Fig.~\ref{mag-consth} (a)), and the cosine of the phase of the magnetization, $\cos \phi_1$ (Fig.~\ref{mag-consth} (b)). 

Let us now, in addition, apply a random field, $\epsilon \vec{\eta}$, in the Y-direction. The new mean field equation is
\begin{eqnarray}
\label{m_h_eps}
\vec{m}=Av_\eta\left[\frac{\int_{0}^{2 \pi} {\vec{\sigma}} \exp(\beta J \vec{m}.\vec{\sigma}+\beta \vec{h}.\vec{\sigma}+\beta \epsilon \eta \sigma_y) 
d\vec{\sigma}}{\int_{0}^{2 \pi}  \exp(\beta J \vec{m}.\vec{\sigma}+\beta \vec{h}.\vec{\sigma}+\beta \epsilon \eta \sigma_y) d\vec{\sigma}}\right],\nonumber \\
\end{eqnarray}
Here we have to solve the two simultaneous equations, given by Eq.~(\ref{m_h_eps}), to obtain the magnitude and the phase of the magnetization vector $\vec{m}$. Just as in the case of a constant field $\vec{h}$ and $\epsilon$ = 0, the solution is unique.
\begin{figure}[t]
\vspace*{+.4cm}
\includegraphics[angle=0,width=70mm]{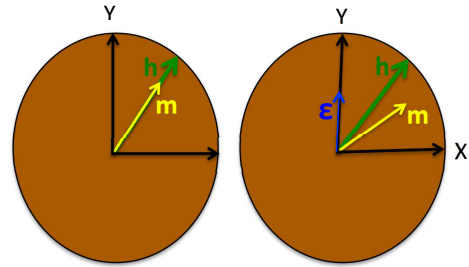}
\vspace*{0.2cm}
\caption{(Color online.) Schematic diagram of the magnetization of XY ferromagnets without and with disorder, in the presence of a constant magnetic field $\vec{h}$. The figure on the left indicates the behavior of $\vec{m}$ in the presence of $\vec{h}$, but when $\epsilon=0$, while the one on the right is when there is a positive $\epsilon$.}
\label{schematic}
\end{figure}

 As in the previous sections, we will now compare the magnetization of the system without disorder (i.e. $\epsilon$ = 0), and for which the mean field equation is given by Eq.~(\ref{m_h})), with the system for which $\epsilon \ne 0$ (and for which the mean field equation is given by Eq.~(\ref{m_h_eps})), keeping $h$ strictly positive in both cases. Let us denote the two Hamiltonians by $H_h$ and $H_{h,\epsilon}$ respectively. We do the comparison by numerical simulations as well as perturbatively at low temperatures (Sec.~\ref{steadyplusrandom-peturb} below). A perturbation approach, similar to the one in	 Sec.~\ref{scaling-low-2d}, can be done at high temperatures also. We refrain from doing it, as the high temperature behavior in this case is less interesting, in view of absence of a phase transition.

The length $m$ of the magnetization vector is shrunk in the system with the disordered field, compared to the ordered case. This is seen from numerical simulations (see Figs.~\ref{mag-consth} (a)), as well as by perturbation techniques at low temperatures. In addition, numerical simulations (shown in Fig.~\ref{mag-consth} (b)) show that the cosine of the phase of the magnetization, i.e., $\cos(\phi_1)$, increases at low temperature in presence of the random field. Therefore, the magnetization vector moves towards the $X$-direction (i.e. the direction transverse to the applied random field). However, the shift turns out to be zero when $x$ equals to $0$, $\pm \pi/2$. This is also corroborated by perturbative analysis at low temperatures. The schematic diagram in Fig.~\ref{schematic} shows the low temperature behavior of the length and phase of the magnetization with and without disorder, in the presence of a constant field. 

The $Y$-component, $m_y = m \sin \phi_1$, of the magnetization has the same relative behavior as the length $m$, in systems described by $H_h$ and $H_{h,\epsilon}$, i.e., for small $\epsilon > 0$ it is lower than for $\epsilon = 0$. 

\begin{figure}[]
\vspace*{+.4cm}
\includegraphics[angle=0,width=70mm]{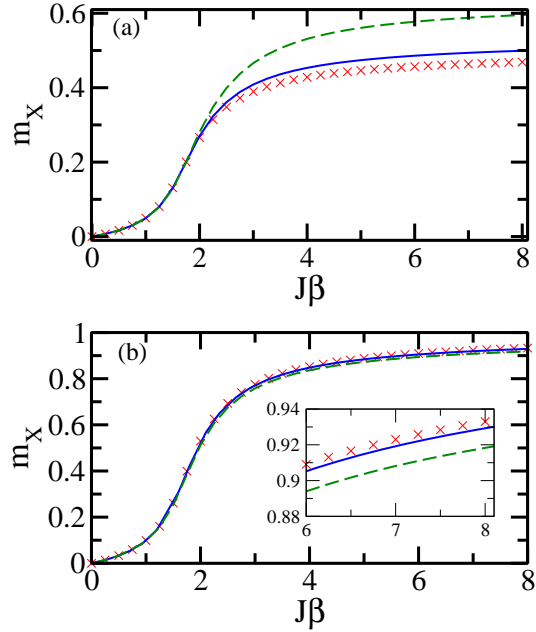}
\vspace*{0.2cm}
\caption{(Color online.) (a) The $X$-component of magnetization, $m_x$, as a function of $J \beta$. Red crosses represent the case when the XY model has the applied constant field $\vec{h}$ with $h/J=0.1$, and (a) $x = \pi/3$ and (b) $x=0.1$. The blue solid and the green dashed lines are for the same system but with the additional random field of strength $\epsilon = 0.1 J$ and $\epsilon = 0.2 J$, respectively. $x$ is measured in radians. All other quantities are dimensionless.}
\label{xcomp_consth}
\end{figure}

However, the $X$-component, $m_x = m \cos \phi_1$, of the magnetization, $\vec{m}$, behaves in a very interesting way. Its value in the system described by $H_{h,\epsilon}$ can be both higher and lower than its value in the system described by $H_h$ depending on the direction of the constant magnetic field. The numerical data for the $X$-component of the magnetic field, $m_x$, are shown for $h/J=0.1$, and $x = \pi/3$ (Fig.~\ref{xcomp_consth} (a)) and $x = 0.1$ (Fig.~\ref{xcomp_consth} (b)). The numerical results for $x=\pi/3$, where $m_x$ is significantly enhanced in presence of the disorder field, signal a random-field induced order: ``order from disorder". However, such effect is absent in the system when $x$ is small (see Figs.~\ref{xcomp_consth} (b) and \ref{PandQ} (top)). We also find that for a given $x$, the shift in $m_x$ due to disorder decreases as the ratio $J/h$ increases.

\subsection{Magnetization at low temperature: Perturbative approach}
\label{steadyplusrandom-peturb}
To obtain the behavior of magnetization at low temperature, we will use the implicit function theorem, which we now state. Let an equation $f(x_1,x_2)=0$ of two variables $x_1$ and $x_2$ be such that $f(x_1,x_2) = 0$ at $(x_1, x_2)=(x^0_1, x^0_2)$. $x_2$ is in	 general an unknown function of $x_1$. But we may still understand the character of $\frac{dx_2}{dx_1} \big|_{x_1=x_1^0}$, by using the fact that (under certain regularity conditions on $f$ near $(x^0_1,x^0_2))$
\begin{equation}
\label{implicit_fun}
\frac{\partial f}{\partial x_1}\biggr{|}_{(x_1^0,x_2^0)}+\frac{\partial f}{\partial x_2}\biggr|_{(x_1^0,x_2^0)} \frac{d x_2}{d x_1}\biggr|_{(x_1^0,x_2^0)} =0.
\end{equation}
The usual statement of the implicit function theorem is that when $\frac{\partial f}{\partial x_2}$ is nonzero at $(x^0_1, x^0_2)$, we can solve the equation $f(x_1,x_2)=0$ for $x_2$ uniquely near this point and the derivative of the resulting function ($x_2$ as a function of $x_1$) at $x_0$ can then be calculated from the above equation. However, in the case when the first derivatives vanish at a certain point, we can use a simple extension of it to calculate the second derivatives. Such a situation appears in the calculations below of the second derivatives of the magnetization with respect to $\epsilon$.

The mean field equations that we work with here can be written in the form
\begin{equation}
\label{meanfield-alternate}
\vec{m}=\frac{1}{\beta J} \triangledown_{\vec{m}} \Gamma,
\end{equation}
where
\begin{equation}
\label{delta}
\Gamma \equiv \log \int \exp(-\beta H_h) \; \text{or} \; \log \int \exp(-\beta H_{h,\epsilon}),
\end{equation}
where $\log$ denotes the natural logarithm.
It follows from symmetry of the distribution of $\eta$ that $\vec{m}$ is an even function of $\epsilon$ and, consequently $\frac{d m_x}{d \epsilon}$ and $\frac{d m_y}{d \epsilon}$ vanish at $\epsilon = 0$. 

\begin{figure}[t]
\vspace*{+.2cm}
\includegraphics[angle=0,width=60mm]{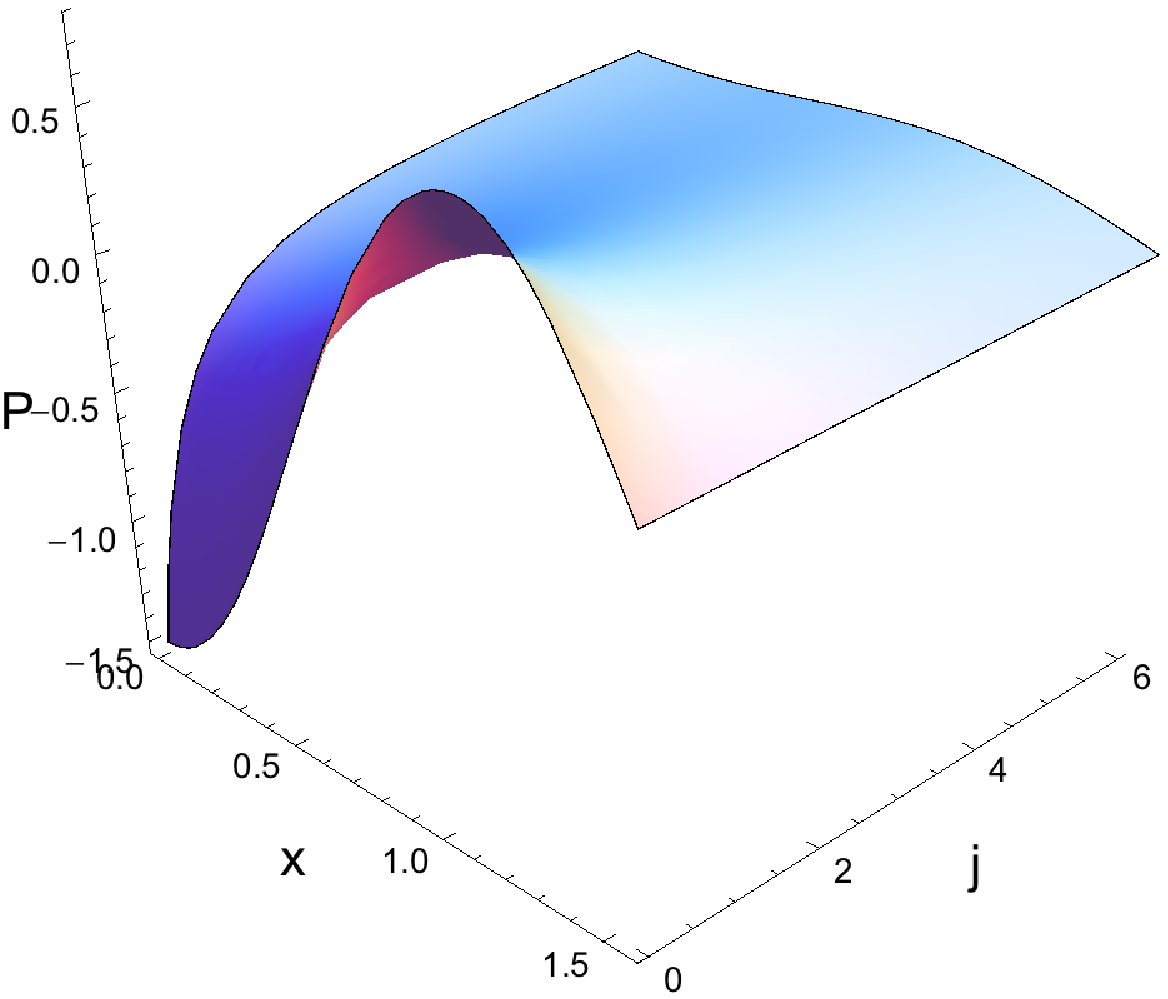}
\includegraphics[angle=0,width=60mm]{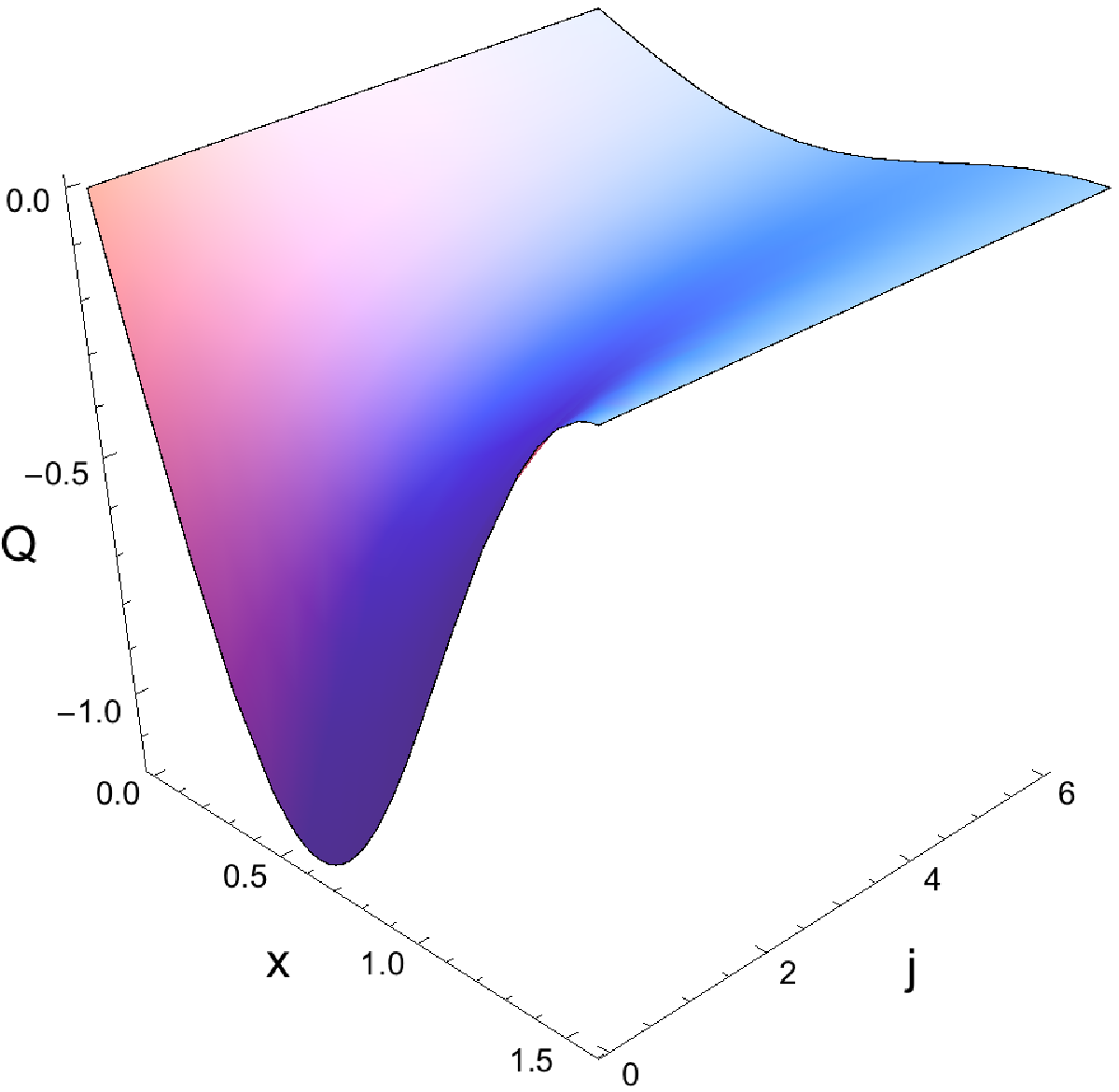}
\vspace*{0.2cm}
\caption{(Color online.) Plot of the functions $P(x,j)$ (top) and $Q(x,j)$ (bottom) with respect to $x$ and $j = J/h$. Note that there are ranges of the $(x,j)$, for which the function $P$ is positive. This fact gives rise to the phenomenon of random field induced order in the system described by the Hamiltonian $H_{h,\epsilon}$. However, $Q(x,j)$ is negative for the entire range of $x$ and $j$. $x$ is measured in radians. All other quantities are dimensionless.}
\label{PandQ}
\end{figure}
It follows that
\begin{equation}
\label{dsqmxy}
\nonumber 
\frac{d^2m_x}{d\epsilon^2}\left[1-\frac{1}{\beta J} \frac{\partial^2 \Gamma}{\partial m_x^2}\right]=\frac{1}{\beta J} \left[ \frac{\partial^3\Gamma}{\partial^2\epsilon\partial m_x}+\frac{\partial^2\Gamma}{\partial m_y \partial m_x} \frac{d^2m_y}{d \epsilon^2}\right]
\end{equation}
and 
\begin{equation}
\frac{d^2m_y}{d\epsilon^2}\left[1-\frac{1}{\beta J} \frac{\partial^2 \Gamma}{\partial m_y^2}\right]=\frac{1}{\beta J} 
\left[ \frac{\partial^3\Gamma}{\partial^2\epsilon\partial m_y}+\frac{\partial^2\Gamma}{\partial m_y \partial m_x} \frac{d^2m_x}{d \epsilon^2}\right],
\end{equation}
where all the total and partial derivatives are taken at $\epsilon = 0$. The above system of equations can be solved for the second (total) derivatives $\frac{d^2m_x}{d\epsilon^2}$ and $\frac{d^2m_y}{d\epsilon^2}$, at $\epsilon=0$, once we can find the partial derivatives at $\epsilon=0$.
The partial derivatives in Eq.~(\ref{dsqmxy}) are calculated by using the following strategy. We have
\begin{equation}
\label{strat-1}
\frac{1}{\beta J} \frac{\partial \Gamma}{\partial m_x}=Av_{\eta} \braket{\cos\theta},
\end{equation}
where for any observable $A$, $\braket{A}$ is the Gibbs average,
\begin{equation}
\label{gibbs-av}
\braket{A}=\frac{\int A \exp(-\beta H)}{\int \exp(-\beta H)},
\end{equation}
with $H$ being the relevant Hamiltonian ($H_h$ or $H_{h,\epsilon}$). Of course, in the case when the system's Hamiltionian is $H_h$, the quenched averaging with respect to $\eta$ is not required. Using this notation we have, differentiating the formula for $\Gamma$ twice,
\begin{equation}
\label{partial-1}
\frac{1}{\beta J}\frac{\partial}{\partial \epsilon} \frac{\partial \Gamma}{\partial m_x}=Av_{\eta}\left[\beta \eta (\braket{\cos \theta \sin \theta}-\braket{\cos \theta}\braket{\sin \theta})\right]
\end{equation}
and
\begin{eqnarray}
\label{partial-2}
\frac{1}{\beta J}\frac{\partial^2}{\partial \epsilon^2} \frac{\partial \Gamma}{\partial m_x}=Av_{\eta}[\beta^2 \eta^2 (\braket{\cos \theta \sin^2 \theta}-
\nonumber\\
2 \braket{\cos \theta \sin \theta} \braket{\sin \theta}+
2 \braket{\cos \theta} \braket{\sin \theta}^2- \braket{\cos \theta}\braket{\sin^2 \theta})]. \nonumber\\
\end{eqnarray}
We expand these partial derivatives with respect to $\frac{1}{\beta}$, at $\epsilon = 0$, using the expansion of the modified Bessel function. After some calculations, we obtain
\begin{equation}
\label{d2mx}
\frac{d^2m_x}{d \epsilon^2}\biggr|_{\epsilon=0}=\frac{1}{h^2} P \left(x,\frac{J}{h} \right) + o(\frac{1}{\beta}),
\end{equation}
and
\begin{equation}
\label{d2my}
\frac{d^2m_y}{d \epsilon^2}\biggr|_{\epsilon=0}=\frac{1}{h^2} Q \left(x,\frac{J}{h} \right) + o(\frac{1}{\beta}),
\end{equation}
where the functions $P$ and $Q$ are given by (for $j = J/h$)
\begin{equation}
\label{P-new}
P (x, j) = \frac{A E+B C}{DE-C^2},
\end{equation}
and
\begin{equation}
\label{Q-new}
Q (x, j) = \frac{A C+B D}{DE-C^2},
\end{equation}
where
\begin{widetext}
\begin{eqnarray}
\label{A-new}
A(x,j)=\frac{1}{(j+1)^3}\big[-\frac{1}{32}((3 j+1)\cos x+(45 j+63) \cos 3x)-
\frac{5}{8}(j+3)\sin 2x \sin x+\frac{1}{4} (j+2) \cos x \cos 2x-
\nonumber\\
\frac{1}{8}(3 j-7) \cos x \sin^2 x\big],
\end{eqnarray}
\begin{eqnarray}
\label{B-new}
B(x,j)=\frac{1}{(j+1)^3}\big[-\frac{1}{32}(3(3 j+1)\sin x+(45 j+63) \sin 3x)-\frac{3}{2}(j+2)\cos 2x \sin x+\frac{3}{8} (j+3) \sin^3 x \big],
\end{eqnarray}
\end{widetext}
\begin{equation}
\label{C-new}
C(x,j) = -\frac{j \cos x \sin x}{j+1},
\end{equation}
\begin{equation}
\label{D-new}
D(x,j) = \frac{j \cos^2 x +1}{j+1},
\end{equation}
\begin{equation}
\label{E-new}
E(x,j) = \frac{j \sin^2 x +1}{j+1}.
\end{equation}
\begin{figure}[t]
\vspace*{+.2cm}
\includegraphics[angle=0,width=60mm]{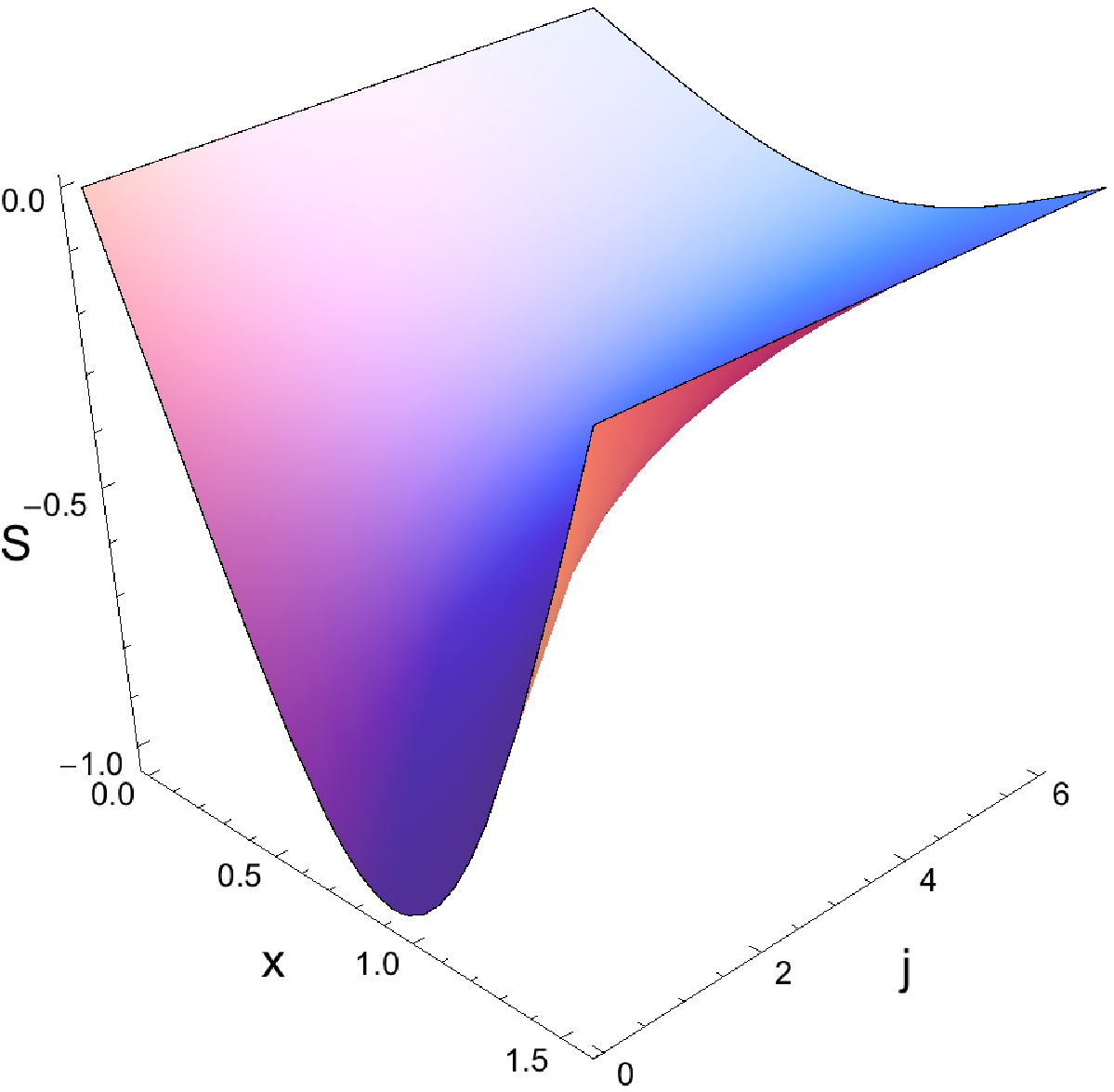}
\includegraphics[angle=0,width=60mm]{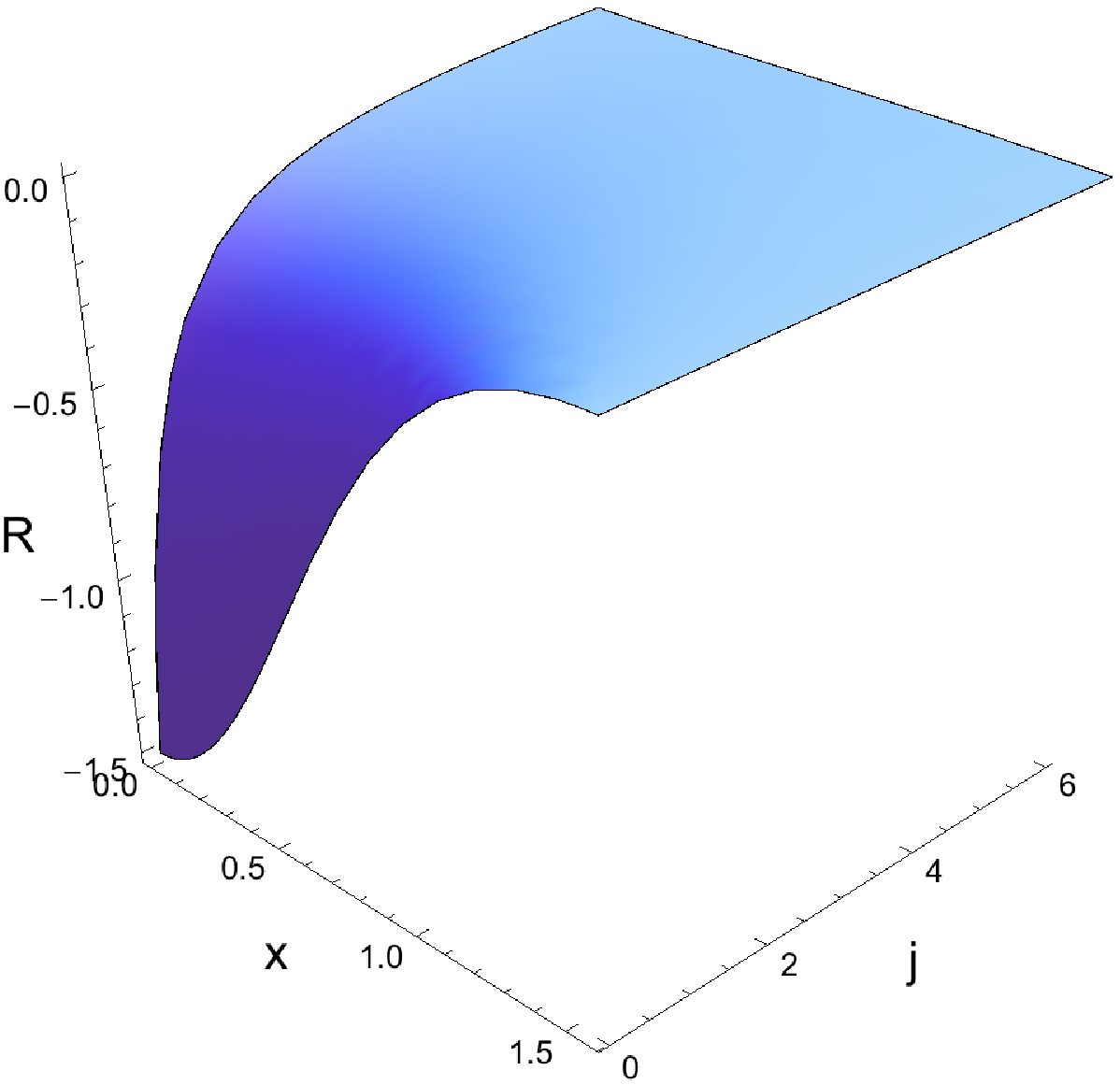}
\vspace*{0.2cm}
\caption{ (Color online.) Plot of the functions $S(x,j)$ (top) and $R(x,j)$ (bottom) with respect to $x$ and $j$. Both of them are again negative for the entire range of $x$ and $j$. These are in agreement with the numerical results in Figs.~\ref{mag-consth}(a) and \ref{mag-consth}(b). $x$ is measured in radians. All other quantities are dimensionless.}
\label{SandR}
\end{figure}
From Fig.~\ref{PandQ} (bottom), it is clear that the $Y$-component of the magnetization always decreases in the presence of disorder. However, Fig.~\ref{PandQ} (top) shows that there are ranges in the parameter space $(x,j)$, for which the quenched averaged $X$-component, $m_x$, of the magnetization increases in the presence of disorder, compared to the case when there is no disorder. As noted before, this is in agreement with our numerical simulations.

We have also considered the effect of disorder on the length $m$ and phase $\phi_1$ of the magnetization. For the phase, we consider the expansion of $\tan(\phi_1) = \frac{m_y}{m_x}$, which is given by
\begin{equation}
\label{phase-1}
\tan(\phi_1)=\frac{m_y}{m_x}\biggr|_{\epsilon=0}+\epsilon^2 \frac{d^2}{d\epsilon^2}\left(\frac{m_y}{m_x}\right)\biggr|_{\epsilon=0}+o(\epsilon^4),
\end{equation}
with
\begin{eqnarray}
\label{phase-2}
 \frac{d^2}{d\epsilon^2}\left(\frac{m_y}{m_x}\right)\biggr|_{\epsilon=0}=\frac{m_x \frac{d^2m_y}{d\epsilon^2}-m_y \frac{d^2m_x}{d\epsilon^2}}{m_x^2}\biggr|_{\epsilon=0}
\nonumber\\
=\frac{1}{m_x^2\big|_{\epsilon=0}}\frac{1}{h^2} S(x,j)+o(\frac{1}{\beta}),
\end{eqnarray}
where
\begin{equation}
\label{S-new}
S(x,j)=Q(x,j) \cos x-P(x,j) \sin x,
\end{equation}
with $P$ and $Q$  given by Eqs.~(\ref{P-new}) and (\ref{Q-new}).

As shown in Fig.~\ref{SandR} (top), $S(x, j)$ is negative for all $x$ and $j$. Consequently, the phase $\phi_1$ always bends towards the $X$-direction in the presence of disorder (since $\tan(\phi_1)$ decreases, cos$(\phi_1)$ increases), as we have already seen in simulations (Fig.~\ref{mag-consth}(b)). Note that $0 \le \phi_1 \le \pi/2$.
The square of the length of the magnetization is given by (up to order $\epsilon^2$)
\begin{equation}
\label{magnitude}
m_x^2+m_y^2=(m_x^2+m_y^2)\big|_{\epsilon=0}+2\epsilon^2\left(R+o\left(\frac{1}{\beta}\right)\right),
\end{equation}
where
\begin{equation}
\label{R-new}
R=(P \cos x+Q \sin x)\big|_{\epsilon=0}.
\end{equation}
As seen in Fig.~\ref{SandR} (bottom), $R$ is always negative, showing that the length of the magnetization decreases in the presence of disorder. Note that the behavior of the length and phase obtained perturbatively, matches with what is shown schematically in Fig.~\ref{schematic}.

\section{Classical Heisenberg model in a random field}
Till now we have explicitly considered only the situation when the spins were two-dimensional. It is natural to ask analogous questions for three-dimensional spins with continuous symmetry. In this section we argue that for the canonical system of this kind - the classical Heisenberg model - the behavior is similar to that of the XY model.

We study the behavior of magnetization in the presence of disorder in the lattice Heisenberg model where the spins are three-dimensional with continuous symmetry. We assume that at all sites, random fields are directed in the Z-direction. The mean-field Hamiltonian of the system is given by Eq.~(\ref{ham_random}). Here, we parameterize $\vec{\sigma}$ as $(\sin \theta_1 \cos \theta_2, \sin \theta_1 \sin \theta_2, \cos \theta_1)$ and we choose $\vec{m}$ 
as $\vec{m}=(m \sin \phi_2 \cos \phi_1, m \sin \phi_2 \sin \phi_1, m \cos \phi_1)$. Therefore, the mean field equation reads
\begin{equation}
\label{h_3d}
\vec{m}=Av_{\eta}\left[\frac{\int \vec{\sigma}\exp(\beta J \vec{m}.\vec{\sigma}+\beta \epsilon \eta \sigma_z) \sin \theta_2 d\theta_2 d\theta_1}{\int \exp(\beta J \vec{m}.\vec{\sigma}+\beta \epsilon \eta \sigma_z) \sin \theta_2 d\theta_2 d\theta_1}\right].
\end{equation}
\subsection{The pure system: $\epsilon=0$}
Consider first the case, when $\epsilon=0$. After some algebra, it can be shown that the three components of the mean field equation (Eq.~(\ref{h_3d})) reduce to the single equation
\begin{equation}
\label{h_3d_f}
-m-\frac{1}{J \beta m}+ \coth(J \beta m)=0.
\end{equation}
Numerical simulations provide the following picture. The pluses in Fig.~\ref{roots_3d}(a) show two real solutions of the Eq.~(\ref{h_3d_f}). A nontrivial solution appears only when $\beta$ is greater than a certain critical temperature $\beta_c^{0,3} \approx 3$ and  approaches unity at very low temperature. The system behaves uniformly in all possible directions of the 3D space, i.e., the solutions of Eq.~(\ref{h_3d}) form a sphere of radius $m_0$ for a given magnetization $\vec{m_0}$.

To find $\beta_c^{0,3}$ analytically, we perform the Taylor series expansion of Eq.~(\ref{h_3d_f}) in $m$ around $m=0$. We obtain
\begin{equation}
\label{approx_3d}
(-1+\frac{J\beta}{3})m-\frac{1}{45}(J^3\beta^3)m^3+o(m^4)=0.
\end{equation}
The nontrivial solutions of Eq.~(\ref{approx_3d}) are given by
\begin{equation}
\label{sol_m0}
m^{0,3} = \pm \frac{\sqrt{15}} {J^{3/2}} \beta^{-3/2} (J\beta-3)^{1/2}.
\end{equation}
$m^{0,3}$ vanishes if $J \beta=3$ and is nonzero  iff  $J \beta \ge 3$, which implies that the critical temperature is given by
\begin{equation}
\label{sol_m01}
\beta_{c}^{0,3}=\frac{3}{J}.
\end{equation}
Note that in the case of the XY model, we also found a similar behavior of the magnetization near its critical temperature.
\subsection{The system with disorder: $\epsilon \ne 0$}
Breaking down $\vec{m}$ in Eq.~(\ref{h_3d}) into its components along the $X$, $Y$ and $Z$ axes, we have three different equations. It turns out that the equations along the $X$ and $Y$ axes, both of which are in a direction transverse to the applied field, reduces to identical equations and we are effectively left with the following two equations to solve:
\begin{flalign}
\label{h_3d_tran}
& m \sin \phi_2 =\nonumber\\
&
Av_{\eta}\left[\frac{\int \sin^2 \theta_2 \exp(\beta (J m \cos \phi_2 +\epsilon \eta ) )I_1[\zeta] d\theta_2}{\int \sin \theta_2 \exp(\beta (J m \cos \phi_2 +\epsilon \eta ) )I_0[\zeta] d\theta_2}\right]
\end{flalign}
and
\begin{flalign}
\label{h_3d_par}
& m \cos \phi_2 = \nonumber\\
&                                  
Av_{\eta}\left[\frac{\int  \sin \theta_2 \cos \theta_2 \exp(\beta (J m \cos \phi_2 +\epsilon \eta ) )I_0[\zeta] d\theta_2}{\int \sin \theta_2 \exp(\beta (J m \cos \phi_2 +\epsilon \eta ) )I_0[\zeta] d\theta_2}\right],
\end{flalign}
where $\zeta=J \beta m \sin \phi_2 \sin \theta_2$.
\begin{figure}[h]
\vspace*{+.1cm}
\includegraphics[angle=0,width=70mm]{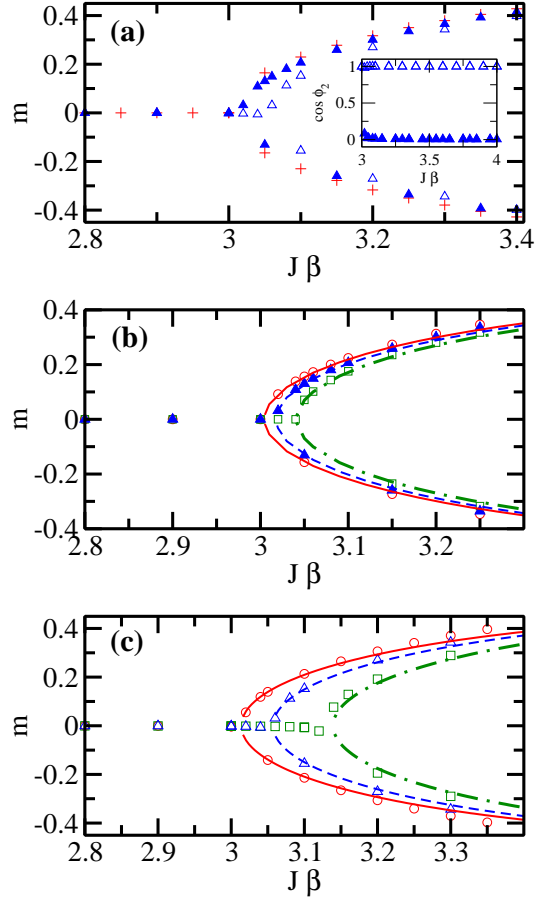}
\vspace*{0.1cm}
\caption{(Color online.) Magnetization as a function of $J\beta$ for the Heisenberg model. (a) Pluses are the roots for the pure system. The filled-in triangles and the empty triangles are the roots for the system with disorder with $\epsilon/J=0.1$ for case I and case II, respectively. Inset: Cosine of the angle associated with the magnetization vector $\vec{m}$ as a function of $J\beta$ for the two different cases. Similar symbols as in the main diagram are used in the inset to represent the two different cases. The data, for which $\cos \phi_2 \approx 1$, suggests that the magnetization belongs to case I. Otherwise, it belongs to case  II, where  $\cos \phi_2 \approx 0$. (b) The circles, triangles and squares are the numerical data for $\epsilon/J $=0.05, 0.1, and 0.15, respectively, for the case I, i.e., for the transverse magnetization. The solid, dashed and dashed-dotted lines correspond to the analytical expression given in Eq.~(\ref{3d_tran_m}) for the same $\epsilon/J$ respectively. (c) The circles, 
triangles and squares are the numerical data for $\epsilon/J $=0.05, 0.1, and 0.15, respectively, for the case II, i.e., for the parallel magnetization. The solid, dashed and dashed-dotted lines correspond to the analytical expression given in Eq.~(\ref{m_parallel_3d}) for the same $\epsilon/J$ respectively. All quantities are dimensionless, except $\phi_2$, which is measured in radians. }
\label{roots_3d}
\end{figure}

Note that both the Eqs.~(\ref{h_3d_tran}) and (\ref{h_3d_par}) are independent of $\phi_1$, implying that the magnetization along the transverse direction of the applied random field forms a circle. A contour analysis, analogous to that done for the XY model, by performing a Taylor series expansion of the equations in $\epsilon$, shows that the behavior of the disordered Heisenberg system is qualitatively similar to that of the XY model with disorder. The system still possesses finite magnetization below a certain critical temperature. The disorder, however, breaks the spherical symmetry and the system possesses magnetization if $\phi_2 = \pi/2$, i.e., along the transverse direction (case I) or if $\phi_2 = 0$, i.e., along the parallel direction (case II) of the applied random field. The numerically obtained solutions of Eqs.~(\ref{h_3d_tran}) and (\ref{h_3d_par}) for the transverse magnetization and the parallel magnetization are shown in Fig.~(\ref{roots_3d})(b) and Fig.~(\ref{roots_3d})(c), respectively. 
For transverse (parallel) magnetization, the system exhibits finite magnetization below a critical temperature, given by $\beta_{\perp,c}^{\epsilon,3}$ $\big(\beta_{\parallel,c}^{\epsilon,3}\big)$. The critical temperature decreases with increasing strength of the randomness. The parallel magnetization remains smaller than that of transverse magnetization in the small $m$ regime (see Fig.~\ref{roots_3d} (a)). 

\subsubsection{Scaling of the transverse magnetization near criticality}
We take advantage of our knowledge about the specific directions of the magnetization that the system can possess. For the transverse magnetization, we put $\phi_2=\pi/2$ in the Eqs.~(\ref{h_3d_tran}) and (\ref{h_3d_par}) and perform a Taylor  expansion of both denominator and numerator in powers of $\epsilon$ and $m$ ($\epsilon$ and $m$ both are small in our regime of interest). Finally, performing the integrations and simplifying the expressions further, Eq.~(\ref{h_3d_par}) becomes trivial and Eq.~(\ref{h_3d_tran}) leads to
\begin{eqnarray}
\label{3d_tran_appox}
((-1+\frac{J\beta}{3})m-\frac{1}{45}J^3 \beta^3m^3+o(m^4))+
\nonumber \\
(-\frac{1}{45}J \beta^3 m+\frac{4}{945} J^3 \beta^5 m^3+o(m^4)) \epsilon^2+o(\epsilon^3)=0. \nonumber\\
\end{eqnarray}
Solving Eq.~(\ref{3d_tran_appox}), we have
\begin{eqnarray}
\label{3d_tran_m}
m_{\perp}^{\epsilon,3}=\pm \frac{\sqrt{21}\sqrt{45-15 J \beta+J\beta^3\epsilon^2}}{\sqrt{-21 J^3\beta^3+4 J^3\beta^5 \epsilon^2}},
\end{eqnarray}
which can again be written as
\begin{eqnarray}
\label{m_perp_3d}
m_{\perp}^{\epsilon,3} \approx \pm m^{0,3} \left(1\mp\frac{\beta^2}{10 (J \beta-3)} \epsilon^2\right). 
\end{eqnarray} 
$\beta_{\perp,c}^{\epsilon,3}$ can then be obtained by solving 
\begin{eqnarray}
\label{3d_crit_con}
45-15 J \beta_{c,\perp}^{\epsilon,3}+J ({\beta_{c,\perp}^{\epsilon,3}})^3\epsilon^2=0.
\end{eqnarray}
The solution of Eq.~(\ref{3d_crit_con}) is given by
\begin{equation}
\label{3d_crit}
\beta_{c,\perp}^{\epsilon,3}=\beta_{c}^{0,3}+\frac{9}{5 J^2} \epsilon^2.
\end{equation}

\subsubsection{Scaling of the parallel magnetization near criticality}
Putting $\phi_2=0$ and proceeding in a similar fashion as in the previous paragraph, Eq.~(\ref{h_3d_tran}) is trivially satisfied and Eq.~(\ref{h_3d_par}) is given by
\begin{eqnarray}
\label{3d_par_appox}
((-1+\frac{J\beta}{3})m-\frac{1}{45}J^3 \beta^3m^3+o(m^4))+
\nonumber \\
(-\frac{1}{15}J \beta^3 m+\frac{4}{149} J^3 \beta^5 m^3+o(m^4)) \epsilon^2+o(\epsilon^3)=0. \nonumber \\
\end{eqnarray}
Solving Eq.~(\ref{3d_tran_appox}) we have
\begin{eqnarray}
\label{m_parallel_3d}
m_{\parallel}^{\epsilon,3}=\pm \frac{\sqrt{21}\sqrt{15-5 J \beta+J\beta^3\epsilon^2}}{\sqrt{-21 J^3\beta^3+20 J^3\beta^5 \epsilon^2}}
\\
\approx \pm m^{0,3} \left(1\mp\frac{3 \beta^2}{10 (J \beta-3)} \epsilon^2\right).
\end{eqnarray}
$\beta_{c,3}^{\epsilon}$ can be obtained by solving the following equation:
\begin{eqnarray}
\label{3d_con_crt_pr}
15-5 J \beta_{c,\parallel}^{\epsilon,3}+J ({\beta_{c,\parallel}^{\epsilon,3}})^3\epsilon^2=0.
\end{eqnarray}
The solution of Eq.~(\ref{3d_con_crt_pr}) is given by
\begin{equation}
\label{3d_crit1}
\beta_{c,\parallel}^{\epsilon,3}=\beta_{c}^{0,3}+\frac{27}{5 J^2} \epsilon^2.
\end{equation}

\section{Generalization of the scalings near criticality for $n$-component $SO(n)$ classical spins}
We consider $SO(n)$ $n$-component classical spins, each of which consist of a radial coordinate of unit length and angular coordinates $\theta_1,\theta_2,\cdots,\theta_{n-1}$, where $\theta_{n-1}$ ranges over $[0,2\pi)$ and all other angles range over $[0,\pi]$. Likewise, the $n$-component magnetization vector $\vec{m}$ consists of a radial coordinate of length $m$ and angular coordinates $\phi_1,\phi_2,\cdots,\phi_{n-1}$. Let $m_1,\cdots,m_n$ be the Cartesian coordinates of $\vec{m}$. $m_i$ can be presented in terms of the radial and angular coordinates as 
\begin{eqnarray}
\label{certasian_m}
m_1=m \cos \phi_1
\nonumber \\
m_2= m \sin \phi_1 \cos \phi_2
\nonumber \\
m_3= m \sin \phi_1 \sin \phi_2 \cos \phi_3
\nonumber\\
\cdots
\nonumber\\
\cdots
\nonumber\\
m_{n-1}=m \sin \phi_1 \cdots \sin \phi_{n-2} \cos \phi_{n-1}
\nonumber\\
m_n=m \sin \phi_1 \cdots \sin \phi_{n-2} \sin \phi_{n-1}
\end{eqnarray}
$\sigma_i$, $i=1,\cdots,n$, the $n$-components of the classical spin $\vec{\sigma}$, can be represented analogously by simply replacing $\phi_j$ by $\theta_j$, $j=1,\cdots,(n-1)$, and substituting $m$ by unity. The volume element $d\vec{\sigma}$ of the $n$-dimensional space is given by
\begin{eqnarray}
\label{volume_nd}
d\vec{\sigma}=\sin^{n-2} \theta_1 \sin^{n-3} \theta_2 \cdots \sin \theta_{n-2}
\nonumber\\
 d\theta_1 d\theta_2 \cdots d\theta_{n-1}.
\end{eqnarray}
We assume the disorder of strength $\epsilon$ to be directed along the $\sigma_1$. We start with the general mean field equation, which can be broken down into a set of $n$ equations, each one of which corresponds to a pair of components  $\{m_i,\sigma_i\}$, where $i=1,\cdots,n$. The equation corresponding to the $i^{th}$ component is given by
\begin{equation}
\label{mf_general}
m_i= F_i(m), 
\end{equation} 
where 
\begin{equation}
\label{f_i_m}
F_i(m)=Av_{\eta}\left[\frac{\int  \sigma_i \exp(\beta J m \alpha+\beta \epsilon \eta \cos \theta_1) d\vec{\sigma}}{\int  \exp(\beta J m \alpha+\beta \epsilon \eta \cos \theta_1) d\vec{\sigma}}\right].
\end{equation}
In Eq.~(\ref{f_i_m}), $\alpha$ is the angle between $\vec{m}$ and $\vec{\sigma}$:
\begin{flalign}
\label{alpha}
& \alpha  =\nonumber\\
&
\big[\cos \theta_{1}\cos \phi_{1}+
\sin \theta_{1}\sin \phi_{1} (\cos \theta_{2}\cos \phi_{2}+\sin \theta_{2}\sin \phi_{2} 
\nonumber\\&
(\cdots + \sin \theta_{n-3} \sin \phi_{n-3} (\cos \theta_{n-2}\cos \phi_{n-2}+
\nonumber\\&
\sin \theta_{n-2}\sin \phi_{n-2}\cos(\theta_{n-1}-\phi_{n-1}))))\big].
\end{flalign}
In order to derive a generalized expression, we envisage that our findings  for $SO(2)$ and $SO(3)$ system would extend to higher dimensions $(SO(n))$, and that the system strictly possess either transverse $(\phi_1=\pi/2)$ or parallel $(\phi_1=0)$ magnetization. We shall see that this is indeed the case.

\subsection{Generalized transverse magnetization near criticality}
Let us first consider the case of transverse magnetization.  We have: $F_1(m) = 0$ and $F_i(m) \ne 0$, where $i=2,\cdots,n$. Because of symmetry, it will be actually sufficient to work with any single $F_i(m)$, where $i=2,\cdots,n$, to derive the generalized small $m$ scaling for the transverse magnetization. We choose to work with the $n^{th}$ component of Eq.~(\ref{mf_general}):
\begin{equation}
\label{mf_nth}
(\Pi_{i=1}^{n-1} \sin \phi_i) m = F_n(m),
\end{equation}
where
\begin{equation}
\label{fn_nth}
 F_n(m)=\frac{A(m)}{B(m)}.
\end{equation}
Here,
\begin{flalign}
\label{am}
A(m)=Av_{\eta}\left[\int (\Pi_{i=1}^{n-1} \sin \theta_i) \exp(\beta J m \alpha+\beta \eta \epsilon \cos \theta_1) d\vec{\sigma}\right],
\end{flalign}
and
\begin{equation}
\label{bm}
B(m)=Av_{\eta}\left[\int\exp(\beta J m \alpha+\beta \eta \epsilon \cos \theta_1) d\vec{\sigma}\right].
\end{equation}
As we are interested in the small $m$ regime, we perform a Taylor series expansion of Eq.~(\ref{mf_nth}) to have 
\begin{equation}
\label{mf_nth_appox}
(\Pi_{i=1}^{n-1} \sin \phi_i) m = F'_n(0) m+\frac{1}{3!}F'''_n(0) m^3.
\end{equation}
In order to evaluate $F'_n(0)$, we need to calculate $A(0), B(0), A'(0)$ and $B'(0)$:
\begin{eqnarray}
\label{fm0}
F'_n(0)=\frac{-A(0)B'(0)+A'(0)B(0)}{B(0)^2}.
\end{eqnarray}
Taylor expansion of Eqs.~(\ref{am}) and (\ref{bm}) in powers of $\epsilon$ up to the second order gives
\begin{flalign}
\label{am_appox}
& A(m)=
\nonumber\\
&
\left[\int (\Pi_{i=1}^{n-1} \sin \theta_i) \exp(\beta J m \alpha)(1+\frac{\beta^2 \epsilon^2}{2} \cos^2 \theta_1) d\vec{\sigma}\right],
\end{flalign}
and 
\begin{flalign}
\label{bm_appox}
& B(m)=
\left[\int (\exp(\beta J m \alpha)(1+\frac{\beta^2 \epsilon^2}{2} \cos^2 \theta_1) d\vec{\sigma}\right].
\end{flalign}
Using Eqs.~(\ref{am_appox}) and (\ref{bm_appox}), we finally obtain the following expressions:
\begin{eqnarray}
\label{a0_appox}
A(0)=0,
\end{eqnarray}
\begin{eqnarray}
\label{b0_appox}
B(0)= (1+\frac{\beta^2 \epsilon^2}{2 n}) \int d\vec{\sigma},
\end{eqnarray}
\begin{eqnarray}
\label{ap0_appox}
A'(0)= \left(\frac{J\beta}{n}+\frac{J \beta^3 \epsilon^2}{2 n (n+2)}\right)  (\Pi_{i=1}^{n-1} \sin \phi_i) \int d\vec{\sigma},\nonumber \\
\end{eqnarray}
and
\begin{eqnarray}
\label{bp0_appox}
B'(0)= 0.
\end{eqnarray}
Plugging Eqs.~(\ref{a0_appox}-\ref{bp0_appox}) into Eq.~(\ref{fm0}) and simplifying further we have
\begin{eqnarray}
\label{fpn0_appox}
F'_n(0)=\left(\frac{J\beta}{n}-\frac{J \beta^3 \epsilon^2}{n^2(n+2)}\right) (\Pi_{i=1}^{n-1} \sin \phi_i).
\end{eqnarray}
The generalized form of $F'''_n(0)$ is given by
\begin{eqnarray}
\label{fp3n0_appox}
F'''_n(0)=3!\left(-\frac{J^3\beta^3}{n^2(n+2)}+\frac{4 J^3 \beta^5 \epsilon^2}{n^3(n+2)(n+4)}\right)
\nonumber\\
 \times(\Pi_{i=1}^{n-1} \sin \phi_i). \nonumber \\
\end{eqnarray}
Using Eqs~(\ref{fpn0_appox}-\ref{fp3n0_appox}) in Eq.~(\ref{mf_nth}) and solving for $m$ we obtain two nontrivial solutions:
\begin{equation}
\label{general_m}
m_{\perp}^{\epsilon,n}=\pm \frac{\sqrt{n(n+4)}\sqrt{n^2(n+2)-n(n+2) J \beta+J \beta^3 \epsilon^2}}{\sqrt{-n(n+4)J^3 \beta^3+4 J^3\beta^5\epsilon^2}},
\end{equation}
which can be written as
\begin{equation}
\label{general_m_appox}
m_{\perp}^{\epsilon,n} \approx m^{0,n} \left(1\mp\frac{\beta^2 }{2(n+2)(J \beta-n)}\epsilon^2\right),
\end{equation}
where
\begin{equation}
\label{general_m_appox1}
m^{0,n}=\pm \frac{\sqrt{n(n+2)}}{J^{\frac{3}{2}}}  \beta^{-3/2}(J\beta-n)^{1/2}.
\end{equation}
Therefore, we find that the decrease of the magnitude of the magnetization due to the random field is of the order of $\epsilon^2$ in all dimensions.

The equation for the critical temperature is:
\begin{equation}
\label{con_crit_gen}
n^2(n+2)-n(n+2) J \beta_{c,\perp}^{\epsilon,n}+J {(\beta_{c,\perp}^{\epsilon,n}})^3 \epsilon^2=0,
\end{equation}
implying
\begin{equation}
\label{crit_gen}
 \beta_{c,\perp}^{\epsilon,n} \approx\frac{n}{J}+\frac{n^2}{J^3(n+2)} \epsilon^2.
\end{equation}
Note that the generalized expressions of the scalings and the critical temperature for the pure system can be obtain by simply putting $\epsilon=0$ in Eqs.~(\ref{general_m}) and (\ref{crit_gen}), respectively. It is also worth mentioning that starting with any other component in Eq.~(\ref{mf_general}) won't alter the results for the scaling near criticality and the critical temperature as long as $n \ge 2$. Hence, the transverse solutions will form an $(n-1)$-dimensional hypersphere. We see that the critical temprature decreases when the dimension increases.

\begin{figure}[t]
\vspace*{+.4cm}
\includegraphics[angle=0,width=60mm]{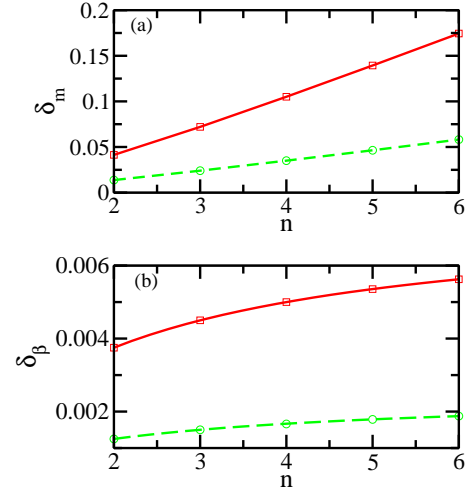}
\vspace*{0.2cm}
\caption{(Color online.) (a) $\delta_m$ and (b) $\delta_\beta$ as functions of the dimension $n$ for the transverse magnetization (green circles) and parallel magnetization (red squares) for the $n$-component classical spin system with $\epsilon/J=0.05$ at $J\beta=J\beta_c+0.1$. The lines serve as guides to the eye. All quantities are dimensionless. Here $\beta_c$ represents $\beta_{c,\perp}^{\epsilon,n}$ for the case of the transverse magnetization and $\beta_{c,\parallel}^{\epsilon,n}$ for the parallel magnetization. }
\label{effdis}
\end{figure}

In order to study the effect of disorder as a function of $n$, we define the dimensionless quantities $\delta_m$ and $\delta_{\beta}$, where
\begin{equation}
\label{delta1}
\delta_m= \bigg|\frac{m^{\epsilon,n}-m^{0,n}}{m^{0,n}}\bigg|
\end{equation}
and
\begin{equation}
\label{delta-beta}
\delta_{\beta}=\frac{\beta_{c}^{\epsilon,n}-\beta_{c}^{0,n}}{\beta_{c}^{0,n}}. 
\end{equation}
$\delta_m$ and $\delta_{\beta}$ are shown in Figs.~(\ref{effdis}) for $\epsilon/J=0.05$.  As the dimension increases, the critical temperature decreases and the disorder becomes effectively stonger. 

\subsection{Generalized parallel magnetization near criticality}
The parallel magnetization can be obtained by setting $F_i(m) = 0$, for $i=2,\cdots,n$:
\begin{equation}
\label{mf_1st}
(\cos \phi_1) m = F_1(m),
\end{equation}
where\begin{equation}
\label{fp_1st}
 F_1(m)=\frac{C(m)}{B(m)}.
\end{equation}
Here,
\begin{equation}
\label{cm}
C(m)=Av_{\eta}\left[\int (\cos \theta_1) \exp(\beta J m \alpha+\eta \epsilon \cos \theta_1) d\vec{\sigma}\right].
\end{equation}
Proceeding as before, we obtain the following expression for the parallel magnetization:
\begin{equation}
\label{general_m_par}
m_{\parallel}^{\epsilon,n}=\pm \frac{\sqrt{n(n+4)}\sqrt{n^2(n+2)-n(n+2) J \beta+3 J \beta^3 \epsilon^2}}{\sqrt{-n(n+4)J^3 \beta^3+20 J^3\beta^5\epsilon^2}},
\end{equation}
which can be written as
\begin{equation}
\label{general_m_appox_par}
m_{\parallel}^{\epsilon,n} \approx m^{0,n} \left(1\mp\frac{3 \beta^2 }{2(n+2)(J \beta-n)}\epsilon^2\right).
\end{equation}
The red squares in Fig.~(\ref{effdis}) show (a) $\delta_m$  and (b) $\delta_\beta$ as a function of $n$ for the parallel magnetization. Both $\delta_m$ and $\delta_\beta$ for the parallel magnetization always remain higher than their analogs for the transverse magnetization in the near-critical regime.  

The equation for the critical temperature is given by
\begin{equation}
\label{con_crit_gen_par}
n^2(n+2)-n(n+2) J \beta_{c,\parallel}^{\epsilon,n}+3 J {(\beta_{c,\parallel}^{\epsilon,n}})^3 \epsilon^2=0,
\end{equation}
which can be solved to obtain
\begin{equation}
\label{crit_gen_par}
 \beta_{c,\parallel}^{\epsilon,n} \approx\frac{n}{J}+\frac{3 n^2}{J^3(n+2)} \epsilon^2.
\end{equation}

\section{Conclusions}
To summarize, this paper consideres classical spin systems within the mean field framework, and studies the effect on magnetization caused by the interplay between a continuous symmetry and a symmetry-breaking quenched disordered field.

We investigated the classical XY and Heisenberg spin systems and showed that even though the symmetry-breaking quenched disordered field destroys the system's continuous symmetry, the system still magnetizes, but only in specific directions, either along the direction of the disordered field or along its transverse. We find that the critical temperatures decreases with increasing strength of the disorder and the magnitude of the magnetization decreases when $\epsilon$ increases.  Moreover, we treated the $n$-component spin model to obtain the near critical scalings of the magnetizations. We found that although the decrease of the magnitude of the magnetization due to the random field, of order $\epsilon$, is of the order of $\epsilon^2$ in all dimensions, the effect of disorder increases with the dimension and consequently the magnetization decreases faster  with increasing dimension when disorder is present. In addition, we studied the classical XY system under the influence of an additional steady field.  
The magnitude of magnetization, $m$, is reduced in the disordered system compared to that of the system without disorder. In the low-temperature regime, the magnetization vector moves towards the transverse direction of the applied random field in presence of the external field.  The disordered system exhibits random field induced ordering in the transverse component of the magnetic field in presence of a uniform magnetic field.

In the future, it will be interesting to adopt mean field approach to investigate the spin systems in random fields in the quantum limit.

\appendix

\section{Derivation of Eq.~(\ref{mnew_pure})}
\label{appendix_symmetry}

Equation (\ref{m_pure}) can be broken down into the following two equations:
\begin{equation}
\label{mx_pure}
m  \cos \phi_1=\frac{\int_{0}^{2 \pi}  \cos \theta_1 \exp(\beta J m \cos(\theta_1-\phi_1)) d\theta_1}{\int_{0}^{2 \pi}  \exp(\beta J m \cos (\theta_1-\phi_1)) d\theta_1},
\end{equation}
and
\begin{equation}
\label{my_pure}
m \sin \phi_1=\frac{\int_{0}^{2 \pi}  \sin \theta_1 \exp(\beta J m \cos (\theta_1-\phi_1)) d\theta_1}{\int_{0}^{2 \pi}  \exp(\beta J m \cos (\theta_1-\phi_1)) d\theta_1},
\end{equation}
where $\theta_1$ and $\phi_1$ are the angles associated with $\vec{\sigma}$ and $\vec{m}$, respectively, i.e., $\vec{\sigma} = (\cos \theta_1, \sin \theta_1)$ and $\vec{m} = (m \cos \phi_1, m \sin \phi_1)$.
Equations (\ref{mx_pure}) and (\ref{my_pure}) both reduces to 
\begin{eqnarray}
\label{m_app_pure}
m = \frac {I_{1} [\beta J m]}{I_{o} [\beta J m]},
\end{eqnarray}

where we have used the following identities:
\begin{equation}
\label{identity-1}
\int_{0}^{2 \pi}  \cos \phi \exp(r  \cos (\phi-t)) d\theta=\cos t\int_{0}^{2 \pi}  \cos \phi \exp(r \cos \phi) d\phi,
\end{equation}
\begin{equation}
\label{identity-2}
\int_{0}^{2 \pi}  \sin \phi \exp(r \cos (\phi-t)) d\theta=\sin t\int_{0}^{2 \pi}  \cos \phi \exp(r \cos \phi) d\phi,
\end{equation}
\begin{equation}
\label{identity-3}
\int_{0}^{2 \pi}  \sin \phi \exp(r \cos \phi) d\phi =0,
\end{equation}
and
\begin{equation}
\label{identity-4}
\int_{0}^{2 \pi}  \cos (n \phi) \exp(r \cos \phi) d\phi = 2 \pi I_n[r],
\end{equation}
and where $I_{n}[x]$ is the modified Bessel function of order $n$ with argument $x$.

\section{Modified Bessel function and its expansion for large arguments}
\label{expand_bessel}

Throughout our work, we had numerous occasions to use the expansion of the modifed Bessel function $I_n(z)$ for large $|z|$ \cite{Abramowitz}. We write down the expression for convenience:
\begin{eqnarray}
\label{bessel}
I_n(z)=\frac{\exp(z)}{\sqrt{2 \pi z}}\big[1-\frac{\mu-1}{8 z}+\frac{(\mu-1)(\mu-9)}{2! (8 z)^2}
\nonumber\\
-\frac{(\mu-1)(\mu-9)(\mu-25)}{3! (8 z)^3}+o(1/z^4)\big],
\end{eqnarray}
where $n$ is fixed and $\mu = 4 n^2$. Actually, the function is well-defined and its expansion  true \cite{Abramowitz} for certain complex ranges of the parameter $z$. However, we will only use them for real $z$.

\end{document}